\documentclass[a4paper,11pt]{article}
\usepackage{jheppub} 

\usepackage{amssymb}
\usepackage{here}
\newcommand{\bel}{\begin{equation}\label}

\newcommand {\beq}{\begin{equation}}
\newcommand {\eeq}{\end{equation}}
\newcommand {\beqa}{\begin{eqnarray}}
\newcommand {\eeqa}{\end{eqnarray}}
\newcommand {\bc}{\begin{center}}
\newcommand {\ec}{\end{center}}
\newcommand {\tr}{{\rm tr\,}}

\newcommand {\vev}  [1]{\ensuremath{\langle #1 \rangle}}
\newcommand {\VEV}  [1]{\ensuremath{\left\langle #1 \right\rangle}}
\newcommand {\rf}   [1]{(\ref{#1})}

\def\dag{\dagger}

\def\vs5{\vspace*{5mm}}
\def\vs1{\vspace*{1cm}}
\def\vs2{\vspace*{2cm}}
\def\hs5{\vspace*{5mm}}
\def\hs1{\hspace*{1cm}}
\def\hs2{\hspace*{2cm}}
\def\vs50{\vspace*{50mm}}
\def\vs20{\vspace*{20mm}}

\def\tr{\hbox{tr}}

\newcommand\ptp[3]   {{\it Prog.\ Theor.\ Phys.\ }{\bf #1} (#2) #3}
\newcommand{\hepth}[1]{\href{http://xxx.lanl.gov/abs/hep-th/#1}{\tt hep-th/#1}}

\preprint{KEK-TH-1634}
\title{Monte Carlo studies of 
the spontaneous rotational symmetry breaking
in dimensionally reduced 
super Yang-Mills models
}
\author[a]{Konstantinos~N.~Anagnostopoulos,}
\author[b]{Takehiro~Azuma}
\author[c,d]{and Jun~Nishimura}
\affiliation[a]{Physics Department, National Technical University,\\
  Zografou Campus, GR-15780 Athens, Greece} 
\affiliation[b]{Institute for Fundamental Sciences, Setsunan
  University,\\
17-8 Ikeda Nakamachi, Neyagawa, Osaka, 572-8508, Japan} 
\affiliation[c]{KEK Theory Center, High Energy Accelerator Research
  Organization,\\
1-1 Oho, Tsukuba, Ibaraki, 305-0801, Japan}
\affiliation[d]{Department of Particle and Nuclear Physics,\\
Graduate University for Advanced Studies (SOKENDAI),\\
1-1 Oho, Tsukuba,
  Ibaraki, 305-0801, Japan} 
\emailAdd{konstant@mail.ntua.gr}
\emailAdd{azuma@mpg.setsunan.ac.jp}
\emailAdd{jnishi@post.kek.jp}
\abstract{
It has long been speculated that
the spontaneous symmetry breaking (SSB) of SO($D$) 
occurs in matrix models
obtained by 
dimensionally reducing super Yang-Mills theory in $D=6,10$ dimensions.
In particular, the $D=10$ case corresponds to the IIB matrix model,
which was proposed as a nonperturbative formulation of superstring theory,
and the SSB may correspond to 
the dynamical generation of four-dimensional space-time.
Recently, it has been shown 
by using the Gaussian expansion method
that the SSB indeed occurs for $D=6$ and $D=10$,
and interesting nature of the SSB common to both cases has been suggested.
Here we study the same issue from first principles
by a Monte Carlo method in the $D=6$ case.
In spite of a severe complex-action problem,
the factorization method enables us to obtain
various quantities associated with the SSB,
which turn out to be consistent with the previous results
obtained by the Gaussian expansion method.
This also demonstrates the usefulness of the factorization method
as a general approach to 
systems with the complex-action problem or the sign problem.
}
\keywords{Matrix Models, Nonperturbative Effects, Spontaneous Symmetry Breaking}

\begin{document} 
\maketitle\flushbottom
\section{Introduction}
\label{sec:intro}
Monte Carlo calculations in lattice gauge theories have played an
important role in nonperturbative studies of QCD. Similar progress
has started to take place in the study of nonperturbative effects in
superstring theory using matrix models. 
(See ref.~\cite{Nishimura:2012xs} for a comprehensive review.) 
Such effects are thought to
play a crucial role in the choice of the true vacuum in the string
landscape. 
The most fundamental problem
is the determination of the macroscopic
space-time dimensionality, which should turn out to be lower than the
space-time dimensionality of the full theory.

The IIB matrix model \cite{Ishibashi:1996xs} has been
conjectured to define superstring theory nonperturbatively in a
properly taken large-$N$ limit of the $N\times N$ matrices. Using
dualities, it is expected that the model describes the unique
underlying theory in spite of its explicit connection to perturbative
type IIB superstring theory. The model is defined originally in 10
dimensions and it has only one scale, even after taking quantum
effects into consideration, which raises the possibility that there is
a unique true vacuum. Space-time in this model
arises dynamically from the eigenvalue distribution of the
ten bosonic matrices $A_\mu$ \cite{9802085}. 
It is therefore
possible to realize dynamical compactification of the
extra dimensions and to obtain a macroscopic four-dimensional
space-time.\footnote{A closely related idea in superstring theory
is to realize an ``emergent space-time'' in the context of
gauge-gravity duality \cite{Maldacena:1997re,Itzhaki:1998dd}. 
To test this idea,
Monte Carlo studies \cite{Hanada:2007ti,Catterall:2007fp,%
Anagnostopoulos:2007fw,%
Catterall:2008yz,Hanada:2008ez,Catterall:2009xn,%
Hanada:2008gy,Hanada:2009ne,Hanada:2011fq}
have been performed on the one-dimensional reduction of the
ten-dimensional ${\cal N}=1$ U($N$) super Yang-Mills theory, which
provides a low energy description of a stack of $N$ D0 branes in type
IIA superstring theory. 
In particular, 
the black hole thermodynamics
has been reproduced from Monte Carlo 
studies of the gauge theory \cite{Hanada:2008ez} 
{\it including $\alpha'$ corrections},
which correspond to the effects of closed strings with finite length.
Wilson loops and correlation functions
of the strongly coupled gauge theory can be calculated more
simply on the gravity side \cite{Rey:1998ik,SY},
and these predictions 
have been confirmed by Monte Carlo simulations on the gauge theory
side \cite{Hanada:2008gy,Hanada:2009ne,Hanada:2011fq}.}

Originally the IIB matrix model has been studied in its
Euclidean version mainly because quantum effects render the partition
function of the model finite despite the flat directions in the action
\cite{Krauth:1998xh,Austing:2001bd}.  Using the Gaussian expansion
method (GEM) \cite{Nishimura:2001sx,higher}, it was shown that
dynamical compactification is realized by spontaneously breaking the
SO(10) rotational symmetry of the model, giving rise to an SO(3)
symmetric vacuum \cite{Nishimura:2011xy}.
Moreover, SO($d$) symmetric vacua in general
were found to have a universal scale
$r$ for the small dimensions
and a scale $R_d$ for the large dimensions satisfying
$(R_d)^d r^{10-d}=\ell^{10}$
with some dynamical scale $\ell$,
resulting in constant volume 
and a finite ratio $R_d/r$.
The scenario of dynamical compactification via SSB was conjectured
earlier based on
the low-energy effective theory \cite{9802085}
and 
the effects
of the fermionic partition function \cite{Nishimura:2000ds,0108041}.

A more exciting scenario has been discovered recently by a Monte Carlo
study of the {\it Lorentzian} version of the IIB matrix model
\cite{Kim:2011cr}. The model was not studied before beyond the
classical level \cite{Steinacker:2011wb,Chatzistavrakidis:2011su}
because it seemed unstable
due to the non-positive-definite bosonic action $S_{\rm b}$ and the phase
factor ${\rm e}^{i S_{\rm b}}$. 
Using simple scaling properties of the model,
it was possible to integrate out ${\rm e}^{i S_{\rm b}}$ yielding a 
constraint $S_{\rm b}\approx 0$. Contrary to the Euclidean model, 
one needs to introduce
large scale cutoffs in the temporal and spatial extents $\frac{1}{N}\tr
(A_0)^2$ and $\frac{1}{N}\tr(A_i)^2 $, which can
be removed in the large-$N$ limit. 
Remarkably, it was found that a 
(3+1)-dimensional {\it expanding} universe
emerges dynamically after a critical time.
Before this
time, space is SO(9) symmetric and small. After this time, 3
dimensions of space expand rapidly signaling the birth of the
universe. Emergence of time happens nontrivially due to a crucial 
role played by supersymmetry, whereas
non-commutativity of space plays an important role in the SSB of SO(9) 
rotational symmetry leading to an SO(3) symmetric space
of large dimensions.
Refs.~\cite{Kim:2011ts,Kim:2012mw} 
investigated classical equations of motion,
which are expected to be valid at late times, 
and presented interesting solutions
which represent a (3+1)-dimensional
expanding universe with zero space-time non-commutativity. 
This result points to the possibility that
space-time non-commutativity disappears at some point in time.
Ref.~\cite{Nishimura:2012rs} has proposed a procedure to identify
the local fields corresponding to the
massless modes that appear at late times.
The possibility that
the Standard Model emerges at low energy
from the IIB matrix model
has been 
discussed in 
refs.~\cite{Aoki:2010gv,Chatzistavrakidis:2011gs,Nishimura:2013moa}.

Monte Carlo studies of the Euclidean IIB and related matrix models,
on the other hand, 
have been pursued for more than fifteen years.
Small matrices were originally studied in
ref.~\cite{Krauth:1998xh} and the large-$N$ limit of 
matrix models without
fermions was studied in ref.~\cite{9811220}. 
The large-$N$ limit of a four-dimensional
version of the IIB
matrix model was first studied in ref.~\cite{Ambjorn:2000bf}.
Simulations of the IIB matrix model and its six-dimensional version
suffer from a strong complex-action
problem. After integrating out the fermionic matrices, the complex
fermionic partition function is found to have a wildly fluctuating
phase.\footnote{The fermionic partition 
function is real
in the Lorentzian IIB matrix model,
which can therefore be studied by Monte Carlo simulation
without the complex-action problem \cite{Kim:2011cr}.}
A phase-quenched model with oneloop approximation 
was studied by Monte Carlo simulation
in ref.~\cite{Ambjorn:2000dx}.
In such models \emph{without} complex action, however,
it is strongly suggested
that SSB does not occur 
\cite{9811220,Ambjorn:2000bf,Ambjorn:2000dx,Ambjorn:2001xs}. 
Therefore, the complex phase is expected to play 
a central role if the SSB really occurs in the Euclidean IIB matrix model.
These early studies,
as well as simulations performed in 
refs.~\cite{Bialas:2000gf,Burda:2000mn}, provided 
a lot of insights into
the large-$N$ limit and the nonperturbative dynamics
of the Euclidean IIB matrix model.

In order to 
overcome the complex-action problem,
ref.~\cite{0108041} proposed the so-called factorization method,
which has the advantage of being quite general.
It was tested on
simple models
\cite{Ambjorn:2002pz,Ambjorn:2004jk,%
Azcoiti:2002vk,Anagnostopoulos:2010ux,Anagnostopoulos:2011cn}
and used in simulations of finite density QCD \cite{Fodor:2007vv} 
(see refs.~\cite{07063549,fdQCD} for related works 
on the QCD phase diagram and refs.~\cite{complex-action-prob} 
for other approaches to the complex-action problem). 
While the actual calculations resemble the density 
of states methods \cite{dos} proposed earlier,
the crucial point 
of the new approach is to
constrain \emph{all} the independent observables that are strongly correlated 
with the phase fluctuations
as was first recognized in refs.~\cite{Anagnostopoulos:2010ux,Anagnostopoulos:2011cn}.
By numerically determining 
the minimum of the free energy with respect to these observables,
it is possible to sample efficiently the configurations 
which give the most important contributions to the partition function.

For the IIB matrix model, the choice of observables
can be made in a rather intuitive way. In ref.~\cite{Nishimura:2000ds} it
is shown that the complex phase vanishes for dimensionally collapsed
configurations and that the phase is stationary with respect to
fluctuations around them. It is therefore possible that 
such configurations
are favored despite their entropic suppression. 
In a simplified matrix model for dynamical compactification
proposed in ref.~\cite{0108070},
we have found strong evidence \cite{Anagnostopoulos:2010ux,Anagnostopoulos:2011cn} 
that the length scales in each dimension are the {\it only}
relevant observables that have an important correlation with the
complex phase. 
Monte Carlo calculations of the simplified model
indeed reproduced various quantities
associated with the SSB obtained earlier by the GEM \cite{0412194}.

In this paper we perform Monte Carlo studies of 
the six-dimensional version of the Euclidean IIB matrix model.
The model is supersymmetric, and the space-time emerges dynamically
from the eigenvalue distribution of the bosonic
matrices. Using the GEM, ref.~\cite{10070883} 
showed that dynamical compactification of the extra dimensions occurs 
via SSB of SO(6) rotational symmetry to an SO(3) symmetric vacuum. 
Similarly to the ten-dimensional model \cite{Nishimura:2011xy},
SO($d$) symmetric vacua in general were found to have
a universal scale $r$
for the small dimensions
and a scale $R_d$ for the large dimensions satisfying
$(R_d)^d r^{6-d} = \ell^6$
with some dynamical scale $\ell$,
resulting in constant volume and a finite
ratio $R_d/r$. 
In ref.~\cite{0108041}, simulations of this model using the
oneloop approximation revealed
the strong effect of
the fluctuations of the phase in generating the scales $r$ and $R_d$.
We will show, however, that this approximation fails to capture the
short distance nonperturbative dynamics of the eigenvalues of the
matrices, which play a crucial role in actual determination
of the scales $r$
and $R_d$. In fact we find a strong, nonperturbative, hard core potential
against the collapse of the eigenvalues that leads to nontrivial
solutions for $r$ in the large-$N$ limit. Simulating the full model,
however, requires ${\cal O}(N^2)$ additional computational effort,
and state-of-the-art algorithms used in lattice QCD simulations with
dynamical fermions have to be used in order to make calculations
feasible. We are able to compute expectation values based on
the factorization method using scaling properties of the
distribution functions in order to extrapolate efficiently to the
region of configuration space favored 
at large $N$. 
These scaling properties are similar to the ones found in the
simplified model \cite{Anagnostopoulos:2010ux,Anagnostopoulos:2011cn}. 
Our final results are consistent with the GEM predictions
for the universal scale $r$ and the constant volume property.
Some preliminary results of this work have been presented in 
a proceeding contribution \cite{Anagnostopoulos:2012ib}.

This paper is organized as follows. In section \ref{sec:model} 
we describe the model and review the previous results
obtained by the GEM.
In section \ref{sec:method} we explain the complex-action problem
and discuss how it can be overcome by
using the factorization method.
In section \ref{sec:simulations} we describe our simulations 
and present our numerical results. 
Section \ref{sec:conclusions} is devoted to
a summary and discussions.
The details of the algorithm used 
in the Monte Carlo simulation are given in appendix \ref{sec:algorithm}.

\section{The model and a brief review of previous results}
\label{sec:model}
The IIB matrix model is formally obtained by the dimensional reduction
of $D=10$, ${\cal N}=1$, SU($N$) super Yang-Mills theory
\cite{Ishibashi:1996xs}. 
After the reduction, the ${\cal N}=1$ supersymmetry is
enhanced to ${\cal N}=2$ supersymmetry, which 
leads to the interpretation of
the eigenvalues of $A_\mu$ as the ten-dimensional space-time
coordinates \cite{Ishibashi:1996xs,9802085}. Therefore, space-time is
generated {\it dynamically}
although it is generically non-commutative 
since dominant configurations in the large-$N$ limit
may consist of non-simultaneously-diagonalizable 
matrices \cite{Ambjorn:2000bf}. 

In this paper we study a $D=6$ version of the
IIB matrix model, whose partition function is given by
\begin{eqnarray}
\label{2.1}
Z & = & \int\, 
dA\,d\psi\,d\bar\psi\, {\rm e}^{-S_{\rm b}-S_{\rm f}}\  ,\nonumber\\
S_{\rm b} &=& -\frac{1}{4 g^2}\tr\, \left[ A_\mu,A_\nu\right]^2\  ,\nonumber\\
S_{\rm f} &=& - \frac{1}{2 g^2}\tr\, 
\left(
 \bar\psi_\alpha(\Gamma_\mu)_{\alpha\beta}
  \left[A_\mu,\psi_\beta\right]
                        \right)
\  .
\end{eqnarray}
The $N\times N$ matrices $A_\mu$ 
($\mu=1,\ldots,6$) are traceless and Hermitian,
while $\psi_\alpha$ and $\bar\psi_\alpha$
($\alpha=1,\ldots,4$) are $N\times N$ traceless matrices with
Grassmannian entries. The scale parameter $g$ can be absorbed
by an appropriate rescaling of the matrices,
and we set $g^2 N=1$ without loss of
generality. Then the actions become
\begin{eqnarray}
\label{2.2}
S_{\rm b} &=& -\frac{1}{4}N \tr\, \left[ A_\mu,A_\nu\right]^2 \ , \\
S_{\rm f} &=& -  \frac{1}{2}N \tr\, \left(
 \bar\psi_\alpha(\Gamma_\mu)_{\alpha\beta}
  \left[A_\mu,\psi_\beta\right]
                        \right)\  .
\label{2.2f}
\end{eqnarray}
The integration measure is given by
\begin{eqnarray}
\label{2.3}
dA &=& 
\prod_{a=1}^{N^2-1}\prod_{\mu=1}^6 \frac{dA^a_\mu}{\sqrt{2\pi}}
\ ,
\nonumber\\
d\psi d\bar\psi &=&
\prod_{a=1}^{N^2-1}\prod_{\alpha=1}^4 
d\psi^a_\alpha d\bar\psi^a_\alpha \ ,
\end{eqnarray}
where $A_\mu=\sum_{a=1}^{N^2-1} A^a_\mu T^a$,
$\psi_\alpha=\sum_{a=1}^{N^2-1}\psi^a_\alpha T^a$ and
$\bar\psi_\alpha=\sum_{a=1}^{N^2-1}\bar\psi^a_\alpha T^a$. The SU($N$)
generators $T^a$ are normalized so that $\tr(T^a
T^b)=\frac{1}{2}\delta^{a b}$. The model has an SO($6$) symmetry,
under which $A_\mu$ transform as a vector and $\psi_\alpha$,
$\bar\psi_\alpha$ as Weyl spinors, respectively. The gamma matrices
after Weyl projection are $4\times 4$ matrices, which we take to be
\begin{eqnarray}
\label{2.4}
\Gamma_1 &=& \sigma_1\otimes\sigma_2\  ,\quad
\Gamma_2  =  \sigma_2\otimes\sigma_2\  ,\quad
\Gamma_3  =  \sigma_3\otimes\sigma_2\  ,
\nonumber\\
\Gamma_4 &=&    {\bf 1}    \otimes\sigma_1\  ,\quad
\Gamma_5  =     {\bf 1}    \otimes\sigma_3\  ,\quad
\Gamma_6  =  i {\bf 1}    \otimes       {\bf 1} \  .
\end{eqnarray}

The model \rf{2.1} is formally obtained by
the dimensional reduction
of the $D=6$, ${\cal N}=1$, SU($N$) super Yang-Mills
theory. 
Similar reductions can be considered for $D=3, 4$
and $10$ dimensions, the last being the IIB matrix model
\cite{Ishibashi:1996xs}. The partition function \rf{2.1} is
potentially non-finite due to the non-compact measure and 
the flat directions $[A_\mu , A_\nu] = 0$ 
in the action. 
Quantum effects\footnote{Naively one might
think that quantum effects are canceled due to supersymmetry.
In fact the flat direction is raised by quantum effects from
fermion zero modes that appear in a diagonal bosonic background
as was first recognized in 
ref.~\cite{9802085}.\label{footnote:flat-direction}}, however, 
render it finite
for $D>3$, a fact that has been checked both 
numerically \cite{Krauth:1998xh} and
analytically \cite{Austing:2001bd}.

The model is studied by first integrating out the fermions, after
which we obtain
\begin{equation}
\label{2.5}
Z = \int\, dA\, {\rm e}^{-S_{\rm b}[A]}\, Z_{\rm f}[A]\  ,
\end{equation}
where $Z_{\rm f}[A]$ represents the fermionic partition function
defined by
\beq
Z_{\rm f}[A] = \int \,d\psi\,d\bar\psi\, {\rm e}^{-S_{\rm f}} 
= \det {\cal  M}[A] \ .
\label{def-Zf}
\eeq
The matrix  ${\cal  M}[A]$ 
is a $4(N^2-1)\times 4(N^2-1)$ matrix, whose explicit form 
is given by eq.~\rf{a.4} in appendix \ref{sec:algorithm}. 
For a generic
configuration, 
$\det {\cal M}[A] = |\det {\cal M}[A]| {\rm e}^{i \Gamma}$ is complex, 
and eq.~(\ref{2.5}) becomes
\begin{equation}
\label{2.6}
Z = \int\, dA\, {\rm e}^{-S_0[A]}\, {\rm e}^{i \Gamma[A]}\  ,
\end{equation}
where
\begin{equation}
\label{2.7}
S_0[A] = S_{\rm b}[A] - \log\left| \det {\cal M}[A]\right|\  .
\end{equation}
The complex phase in eq.~\rf{2.6} is a common feature of the ${D=6,10}$
models, whereas for $D=4$, we have $\Gamma[A]\equiv 0$. 
In the latter case, no SSB
occurs \cite{Ambjorn:2000bf,Ambjorn:2001xs}, 
consistent with the observation that the phase plays a central
role in the SSB mechanism. 

The properties of the phase relevant to the SSB were found 
in ref.~\cite{Nishimura:2000ds}, 
and numerical studies in ref.~\cite{0108041}
confirmed the picture by showing that strong fluctuations of the phase
are the main effect that suppresses symmetric configurations and 
favors non-symmetric ones despite their entropic suppression. First
one notes that (i)~configurations with $A_6=0$ give
real\footnote{In this case, the determinant 
is not necessarily positive,
but it turns out that configurations with positive determinant
dominate at large $N$.
\label{footnote:dim-five}
} 
$\det{\cal  M}[A]$, 
(ii)~configurations with $A_5=A_6=0$ give $\det{\cal
  M}[A]\ge 0$,  and 
(iii)~configurations with $A_3=\ldots=A_6=0$ give
$\det{\cal M}[A]\equiv 0$. 
Let us define ``$d$-dimensional configurations''
($1\le d \le 5$) as configurations of $A_\mu$ that can be transformed to
$A_{d+1}=\ldots=A_6=0$ by an appropriate SO($6$) transformation. 
Then, for $d \le 4$, we have 
\begin{equation}
\label{2.8}
\frac
{\partial^k\Gamma[A]}
{\partial A_{\mu_1}\ldots\partial A_{\mu_k}} = 0
\qquad
\mbox{for}
\qquad
k=1,\ldots,6-(d+1) \ .
\end{equation}
This is because up to $6-(d+1)$-th order of perturbation,
the configuration remains $5$-dimensional, and therefore
$\det{\cal M}[A]$ remains real non-negative.
Strictly speaking, eq.~(\ref{2.8}) does not hold
for $d\le 2$
since $\det{\cal M}[A]\equiv 0$ for such configurations,
and the phase $\Gamma[A]$ becomes ill-defined.

In order to probe the SSB of SO($6$) rotational symmetry,
we study the ``moment of inertia'' tensor
\begin{equation}
\label{2.9}
T_{\mu\nu}=\frac{1}{N} \tr\left(A_\mu A_\nu\right)
\end{equation}
and its real positive eigenvalues
$\lambda_n$ ($n=1,\ldots,6$) ordered as
\begin{equation}
\label{2.10}
\lambda_1\ge\lambda_2\ge\ldots\ge\lambda_6\  .
\end{equation}
The vacuum expectation values (VEVs) $\vev{\lambda_n}$, taken {\it after}
the ordering for each configuration, play the role of order
parameters. If they turn out to be unequal in the large-$N$ limit, it
implies SSB of SO($6$).

The VEVs $\vev{\lambda_n}$ have
been calculated in the large-$N$ limit
by the GEM \cite{10070883}
for the SO($d$) symmetric vacuum ($2 \le d \le 5$),
which has
\beq 
\vev{\lambda_{1}}_{{\rm SO}(d)}
= \ldots = \vev{\lambda_{d}}_{{\rm SO}(d)}  \equiv (R_d)^2
\label{Rd-def}
\eeq
due to the ${\rm SO}(d)$ symmetry.
The small eigenvalues 
$\vev{\lambda_{k}}_{{\rm SO}(d)}$ with $k > d$ 
have little dependence on $k$ for each $d$,
and moreover the mean value turns out to be 
universal for all $d$.
The results for $(R_d)^2$ ($3\le d \le 5$), on the other hand,
are fitted nicely 
to\footnote{The GEM results
for $d=2$ do not satisfy this property, which
may be due to the subtlety in the calculations
for the $d=2$ case explained in ref.~\cite{10070883}.}
\begin{equation}
\label{1.1}
(R_d)^d r^{6-d} = \ell^6
\end{equation}
with $r^2\approx 0.223$ and $\ell^2 \approx 0.627$.
The value of $r^2$ turns out to be consistent with 
the universal scale in the small directions.
Therefore, the left-hand side of (\ref{1.1}) actually
represent the six-dimensional
volume of the dynamical space-time, and hence eq.~(\ref{1.1}) is called
the constant volume property.
This implies that the dynamical space-time in this model behaves effectively 
as an incompressible fluid
and that the phase of the fermion determinant can only make it collapsed
without changing its volume. 
Physical understanding of the universal scale $r$ 
and the constant volume property is
discussed in section 6 of ref.~\cite{10070883}
based on low-energy effective theory \cite{9802085}.


\section{The complex-action problem and the factorization method}
\label{sec:method}

Monte Carlo calculations of  $\vev{\lambda_n}$ are quite hard due to
the strong fluctuations of the phase $\Gamma[A]$ 
in eq.~(\ref{2.6}). A straightforward
approach is to simulate the phase-quenched model
\begin{equation}
\label{2.11}
Z_0 = \int \, dA\, {\rm e}^{-S_0[A]}\  ,
\end{equation}
and to compute VEVs
in the full model by
reweighting\footnote{In the second equality, we have used the fact
that in the phase-quenched model, the phase $\Gamma$ flips sign 
under the parity transformation $A_6\to -A_6$. 
Similar remarks apply to eqs.~(\ref{2.17}), (\ref{2.19})
and (\ref{2.28}) as well.} 
\begin{equation}
\label{2.12}
\vev{\lambda_n} = \frac{\vev{\lambda_n\, {\rm e}^{i\Gamma}}_0}
                       { \vev{{\rm e}^{i\Gamma}}_0} 
                = \frac{\vev{\lambda_n\, \cos{\Gamma}}_0}
                       { \vev{\cos{\Gamma}}_0}
\  ,
\end{equation}
where $\vev{ \ \cdot \ }_0$ are 
VEVs
taken with respect to the
phase-quenched model in eq.~\rf{2.11}. This approach suffers from the
complex-action problem and the overlap problem. The VEVs 
$\vev{ \  \cdot \ {\rm e}^{i\Gamma}}_0$ 
decrease exponentially at large $N$ as
${\rm e}^{-N^2 \Delta F}$, where $\Delta F>0$ is the difference of the
free energies of the full and phase-quenched models defined by the
ratio $Z/Z_0$. This happens via huge cancellations 
due to the oscillating terms from the phase factor ${\rm e}^{i \Gamma}$. 
As a result, one needs ${\cal O}({\rm  e}^{\mbox{\scriptsize const.}\times N^2})$ 
configurations to compute an observable with given accuracy. This is
the complex-action problem or the sign problem. 
The overlap problem is due to the
exponentially small overlap of the distribution of the configurations
sampled in $Z_0$ with the important configurations in $Z$ with
increasing $N$.

In order to overcome the complex-action problem and the overlap problem, 
a new method termed the
factorization method was proposed in
refs.~\cite{0108041,Anagnostopoulos:2010ux}. 
We review the refined version \cite{Anagnostopoulos:2010ux}
as applied to the present model.

In what follows,
we study the normalized eigenvalues
\begin{equation}
\label{2.13}
\tilde\lambda_n = \frac{\lambda_n}{\vev{\lambda_n}_0}\  .
\end{equation}
The deviation of $\vev{\tilde\lambda_n}$ from $1$ represents the
effect of the phase. We consider the distribution functions
\begin{eqnarray}
\label{2.14}
\rho(x_1,\ldots,x_6) &=& 
\VEV{\prod_{k=1}^6\delta(x_k-\tilde\lambda_k)}
\\
\label{2.15}
\mbox{and~~~~~}
\rho^{(0)}(x_1,\ldots,x_6) &=& 
\VEV{\prod_{k=1}^6\delta(x_k-\tilde\lambda_k)}_0
\end{eqnarray}
for the full model and the phase-quenched model, respectively. These
functions vanish due to the ordering (\ref{2.10}) 
unless\footnote{In the large-$N$ limit, $\vev{\lambda_n}_0$ approaches
a constant independent of $n$ (see figure \ref{f:01}),
which implies that the condition becomes
$x_1\ge \ldots \ge x_6$.} 
$x_1 \vev{\lambda_1}_0 \ge \ldots \ge x_6 \vev{\lambda_6}_0$. 
By applying the reweighting formula like \rf{2.12}
to the right-hand side of eq.~\rf{2.14}, one finds that it factorizes as
\begin{equation}
\label{2.16}
\rho(x_1,\ldots,x_6)=
\frac{1}{C} \, \rho^{(0)}(x_1,\ldots,x_6) \, w(x_1,\ldots,x_6)\  .
\end{equation}
The function $w(x_1,\ldots,x_6)$ is defined by\footnote{Numerically,
this function is found to be positive, which simplifies our analysis 
considerably. See ref.~\cite{Ambjorn:2002pz} for an analysis 
of a system in which the corresponding function is complex.}
\begin{equation}
\label{2.17}
w(x_1,\ldots,x_6) = \vev{{\rm e}^{i\Gamma}}_{x_1,\ldots,x_6}
= \vev{{\rm cos}\Gamma}_{x_1,\ldots,x_6}
\  ,
\end{equation}
where $\vev{ \ \cdot \ }_{x_1,\ldots,x_6}$ 
denotes a VEV with respect to the
partition function
\begin{equation}
\label{2.18}
Z_{x_1,\ldots,x_6}=\int\, dA\,
 {\rm e}^{-S_0[A]}\,
 \prod_{k=1}^6\delta(x_k-\tilde\lambda_k)\  .
\end{equation}
The real parameter $C$ is a normalization constant given by
\begin{equation}
\label{2.19}
C = \vev{{\rm e}^{i\Gamma}}_0
= \vev{{\rm cos}\Gamma}_0
\  ,
\end{equation}
which is {\it not} needed in the calculations using the factorization
method. 

The VEVs $\vev{\tilde\lambda_n}$ can be written in terms of the
distribution function as
\begin{equation}
\label{2.20}
\vev{\tilde\lambda_n}= \int \prod_{k=1}^6dx_k\, x_n\,\rho(x_1,\ldots,x_6)\  .
\end{equation}
In the large-$N$ limit, the integral is dominated by
the minimum of the ``free energy''
\begin{eqnarray}
\label{free-energy-def}
{\cal F}(x_1, \ldots , x_6) 
&=& -\frac{1}{N^2} \log\rho(x_1, \ldots , x_6)  \\
&=& -\frac{1}{N^2} \log\rho^{(0)}(x_1, \ldots , x_6) 
-\frac{1}{N^2} \log w(x_1, \ldots , x_6) + \frac{1}{N^2} \log C  \ .
\nonumber
\end{eqnarray}
In order to obtain the minimum, we solve 
a set of coupled equations
\begin{equation}
\label{2.21}
\frac{1}{N^2} \frac{\partial}{\partial x_n} 
\log \rho^{(0)}(x_1,\ldots,x_6) =
-\frac{\partial}{\partial x_n} 
\frac{1}{N^2}\log w(x_1,\ldots,x_6)
\quad \mbox{for}\quad
n=1,\ldots,6 \  ,
\end{equation}
where the function on each side has a definite large-$N$ limit.
In fact there are more than one solutions,
and 
we need to identify the minimum eventually by comparing
the free energy at each solution.
This way we can get a robust estimate for the VEVs
$\vev{\tilde\lambda_n}$, which becomes exact in the large-$N$ limit.
Although calculations of 
the right-hand side of eq.~\rf{2.21}
still suffer from the fluctuations of the phase, the effect is
greatly reduced for two reasons: For given $N$, the system is
constrained in the region of configuration space favored by the competition
of entropic effects, the real action and the phase fluctuations. In
this region, phase fluctuations are greatly reduced compared to the
region mainly sampled by the phase-quenched model. Furthermore,
it is possible to extrapolate 
$\dfrac{1}{N^2}\log w(x_1,\ldots,x_6)$ 
to larger values of $N$ than allowed by direct
simulation \cite{0108041}. 

In applications to general complex-action systems,
finding the operators that are strongly correlated with the phase 
is crucial 
for the success of the 
method \cite{Anagnostopoulos:2010ux,Anagnostopoulos:2011cn}.
The choice of the observables 
in the present model
is due to the strong correlation of $\lambda_n$ 
with the phase $\Gamma$ as expected
from the arguments below eq.~\rf{2.7}.
By solving
eq.~\rf{2.21}, one can 
determine the important configurations in eq.~\rf{2.6}.
It is then straightforward to do effective importance sampling 
by constraining the system in the region of the solutions.
Thus the overlap problem is solved.
The VEV of any other observable
weakly correlated with the phase ${\rm  e}^{i\Gamma}$
can be obtained by estimating them
in the constrained system (\ref{2.18})
at the solutions. 
%

In practice, numerically solving eq.~\rf{2.21} 
in its full generality in the 6d
parameter space $(x_1,\ldots,x_6)$ is a formidable 
task.
In a simplified model
studied in refs.~\cite{Anagnostopoulos:2010ux,Anagnostopoulos:2011cn},
we reduced the task by
assuming that some subgroup of the rotational symmetry is
unbroken in each of possible vacua.
Similarly, we would like to study the SO($d$) symmetric vacua with
$2 \le d \le 5$, which correspond to the solutions
to eq.~\rf{2.21} with $x_1 = \ldots = x_d > 1 > x_{d+1}, \ldots , x_6$.
In fact we can reduce the computational task further
by noting that the effect of the phase 
in the present model is such
that some of the eigenvalues $\lambda_{n}$ become maximally small,
and the others become quite large.
The large eigenvalues, as far as they are sufficiently large,
do not affect much the fluctuation of the phase 
as we will see later.
Therefore we may omit the large eigenvalues from the set of observables
to be constrained in the factorization method.
The small eigenvalues, on the other hand,
tend to acquire the same value\footnote{This property has been 
observed in the GEM calculations \cite{10070883}
as we mentioned below eq.~(\ref{Rd-def}).}
for entropic reasons.
This allows us to constrain only $\lambda_{d+1}$,
the largest eigenvalue among the small ones,
when we study the SO($d$) symmetric vacuum.\footnote{After this
simplification, our task reduces \emph{formally}
to that of the single-observable
factorization method, as originally proposed in
ref.~\cite{0108041}. However, we emphasize that the interpretation
we adopt in this work is based on the multi-observable
factorization method \cite{Anagnostopoulos:2010ux}
described above, and it is 
different from the one in ref.~\cite{0108041}.
For instance, the overlap problem may occur when we
calculate (\ref{2.28}) by simulating the system (\ref{2.27})
as pointed out in ref.~\cite{Anagnostopoulos:2011cn}.
This is not a problem, though, when we interpret the solution
to (\ref{2.30}) in the $x<1$ region as an estimate of 
$\vev{\tilde\lambda_{d+1}}_{{\rm SO}(d)}$
for $n=d+1$.
%
}

%
Below we describe how we can estimate the VEVs
$\vev{\tilde\lambda_{n}}_{{\rm SO}(d)}$ and
the free energy ${\cal F}_{{\rm SO}(d)}$
for the SO($d$) symmetric vacuum under the above assumptions.
We define 
\begin{equation}
\label{2.28}
w_n(x) = \vev{{\rm e}^{i \Gamma}}_{n,x} 
= \vev{{\rm cos} \Gamma}_{n,x} 
\ ,
\end{equation}
where the VEV is taken with respect to 
the $\tilde\lambda_n$-constrained system 
\begin{equation}
\label{2.27}
Z_{n,x} = \int d A\, {\rm e}^{-S_0[A]}\,
               \delta(x-\tilde\lambda_n)  \ .
\end{equation}
We also define
\begin{equation}
\label{def-rho-0}
\rho_n^{(0)}(x)=\VEV{\delta(x-\tilde\lambda_n)}_0 \ .
\end{equation}
Under the present assumptions, the problem reduces to
finding a solution $\bar{x}_n$ to
\begin{equation}
\label{2.30}
 \frac{1}{N^2}f^{(0)}_n(x) \equiv
 \frac{1}{N^2}\frac{d}{dx}\log\rho^{(0)}_n(x) = 
- \frac{d}{dx} \frac{1}{N^2} \log w_n(x)\  
\end{equation}
in the $x<1$ region.
Then the solution $\bar{x}_n$ with $n=d+1$ gives an estimate for 
the small eigenvalue $\vev{\tilde\lambda_{d+1}}_{{\rm SO}(d)}$ 
in the SO($d$) symmetric vacuum 
in the large-$N$ limit.
The other eigenvalues in the SO($d$) symmetric vacuum 
can be estimated as
\begin{equation}
\vev{\lambda_{k}}_{{\rm SO}(d)}
= \vev{\lambda_{k}}_{n,\bar{x}_n}
 \ , \quad \mbox{where $n=d+1$} \ .
\label{vev-lambda}
\end{equation}

Free energy for the SO($d$) symmetric vacuum is given by
estimating (\ref{free-energy-def})
at the corresponding solution of eq.~\rf{2.21}.
Under the present assumptions, 
the free energy
for the SO($d$) symmetric vacuum
can be estimated up to a common constant by 
\begin{equation}
{\cal F}_{{\rm SO}(d)} = 
\int _{\bar{x}_n}^1  \frac{1}{N^2}f^{(0)}_n(x)  dx
-\frac{1}{N^2} \log w_n(\bar{x}_n)
 \ , \quad \mbox{where $n=d+1$} \ .
\label{simpler-free-energy}
\end{equation}
By comparing the free energy for different $d$,
we can determine the true vacuum, which gives
the minimum of the free energy (\ref{free-energy-def}).
The other vacua actually correspond to the saddle-points
of the free energy (\ref{free-energy-def}).


\section{Simulations and the results}
\label{sec:simulations}

Monte Carlo simulations are performed on the system
\begin{equation}
\label{3.1}
Z_{n,V} = \int\, dA\, {\rm e}^{-S_0[A] - V(\lambda_n[A])}\  ,
\quad
V(z) = \frac{1}{2}\,\gamma\,(z-\xi)^2\  ,
\end{equation}
where $\gamma$ and $\xi$ are real parameters. For $\gamma$ large
enough, the Gaussian function approximates the delta function in
eq.~\rf{2.27}. We use the 
Rational Hybrid Monte Carlo (RHMC) algorithm \cite{Kennedy:1998cu} 
and the details
of the simulations
are described in appendix \ref{sec:algorithm}.

The distribution functions of $\tilde\lambda_n$ for the system
$Z_{n,V}$ are given by
\begin{equation}
\label{3.2}
\rho_{n,V}(x) \equiv 
\VEV{\delta(x-\tilde\lambda_n)}_{n,V} \propto
\rho^{(0)}_n(x) 
 \exp\left\{-V\left(x\,\vev{\lambda_n}_0\right)\right\}\  ,
\end{equation}
where $\vev{ \ \cdot \ }_{n,V}$ represents a VEV with respect to
$Z_{n,V}$. The position of the peak, which we denote by $x_{\rm p}$, can be
obtained by solving 
\begin{equation}
\label{3.3}
0 = \frac{d}{dx}\log\rho_{n,V}(x) = 
 f^{(0)}_n(x)-\vev{\lambda_n}_0\, V'\left(x\,\vev{\lambda_n}_0\right)\  .
\end{equation}
For sufficiently large $\gamma$, the distribution function
$\rho_{n,V}(x)$ is sharply peaked at $x_{\rm p}$ and we can use the VEV of
$\tilde\lambda_n$ as an estimator for $x_{\rm p}$, i.e.,
\begin{equation}
\label{3.4}
x_{\rm p} = \vev{\tilde\lambda_n}_{n,V}\  .
\end{equation}
By varying the value of $\xi$ in eq.~\rf{3.1}, we obtain the functions
$w_n(x)$ and $f^{(0)}_n(x)$ by
\begin{eqnarray}
w_n(x_{\rm p}) &=& \VEV{\cos\Gamma}_{n,V}\  ,\label{3.5}\\
f^{(0)}_n(x_{\rm p}) &=& \vev{\lambda_n}_0\,
    V'\left(\vev{\lambda_n}_{n,V}\right) =
\gamma \vev{\lambda_n}_0 \left(
  \vev{\lambda_n}_{n,V}-\xi
                        \right)\  .\label{3.6}
\end{eqnarray}
The parameter $\gamma$ should be chosen large enough to make the
fluctuation of $\tilde\lambda_n$ smaller than the required resolution
in $x$. It should not be too large because in that case,
$(\vev{\lambda_n}_{n,V}-\xi)\propto 1/\gamma$, and a small error in
$\vev{\lambda_n}_{n,V}$ propagates to $f^{(0)}_n(x)$ by a factor of
$\gamma$. In practice, one makes sure that the results for a given
value of $x$ are independent of the choice of $(\gamma,\xi)$ up to the
accuracy goal\footnote{Typically we use $\gamma \sim 100-1000$,
but we have tested values up to $\gamma\sim 10^6$ and verified that the
results are independent of $\gamma$.}.
This method of computing 
$f^{(0)}_n(x)$ 
is quite
different from a direct computation of the distribution
$\rho^{(0)}_n(x)$ of $\tilde\lambda_n$ in the phase-quenched model $Z_0$,
where one can obtain reasonable statistics only in the vicinity
of the peak $x=1$. In contrast, our method enables us
to measure in regions of 
configuration space
suppressed by many orders of magnitude.

\begin{figure}[tbp]
\centering 
\includegraphics[width=9cm]{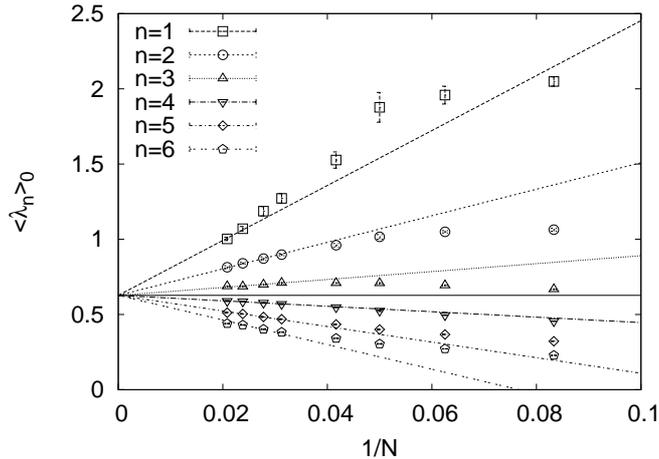}
    \caption{\label{f:01} The eigenvalues 
$\langle \lambda_{n} \rangle_{0}$ for the phase-quenched model
are plotted against $1/N$. 
The solid line represents the value
$\ell^2 = 0.627$, which appears 
in eq.~(\ref{1.1}) obtained by the GEM \cite{10070883}.
The other lines represent the fits to the behavior 
$\vev{\lambda_n}_0 = \ell^2 + {\cal O}(1/N)$.
}
\end{figure}

The phase-quenched model $Z_0$ corresponds to
the special case $\gamma=0$ of eq.~(\ref{3.1}).
We simulate this system and compute $\vev{\lambda_n}_0$, $n=1,\ldots,6$,
the results of which are shown in figure \ref{f:01}.
We find that the data can be nicely fitted to
$\vev{\lambda_n}_0 = c + {\cal O}(1/N)$,
where $c$ is chosen to be 
$\ell^2 = 0.627$,
which appears in the constant volume property (\ref{1.1}) obtained by the GEM.
(This agreement $c=\ell^2$ is consistent 
with the physical interpretation of the 
constant volume property given below eq.~(\ref{1.1}).)
Thus the absence of a fluctuating phase
results in no SSB of SO($6$) as expected
\cite{9811220,Ambjorn:2000bf,Ambjorn:2000dx,Ambjorn:2001xs}.
We should also note that finite-$N$ effects make the eigenvalues
$\vev{\lambda_n}_0$ depend much on $n$.
The normalization (\ref{2.13}) is useful in reducing such finite-$N$
effects and making it easier to see the large-$N$ scaling behavior 
in the following analysis.

\begin{figure}[tbp]
\centering 
\includegraphics[width=7.4cm]{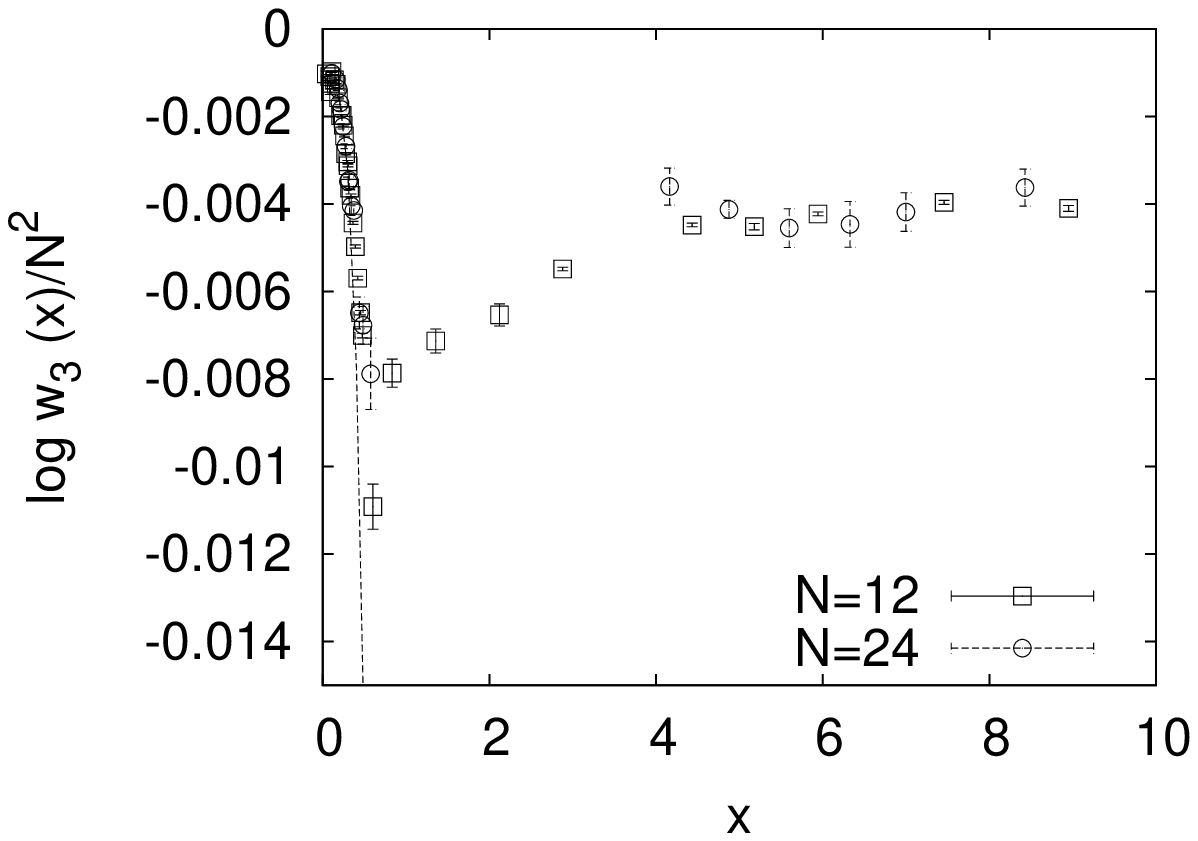}
\includegraphics[width=7.4cm]{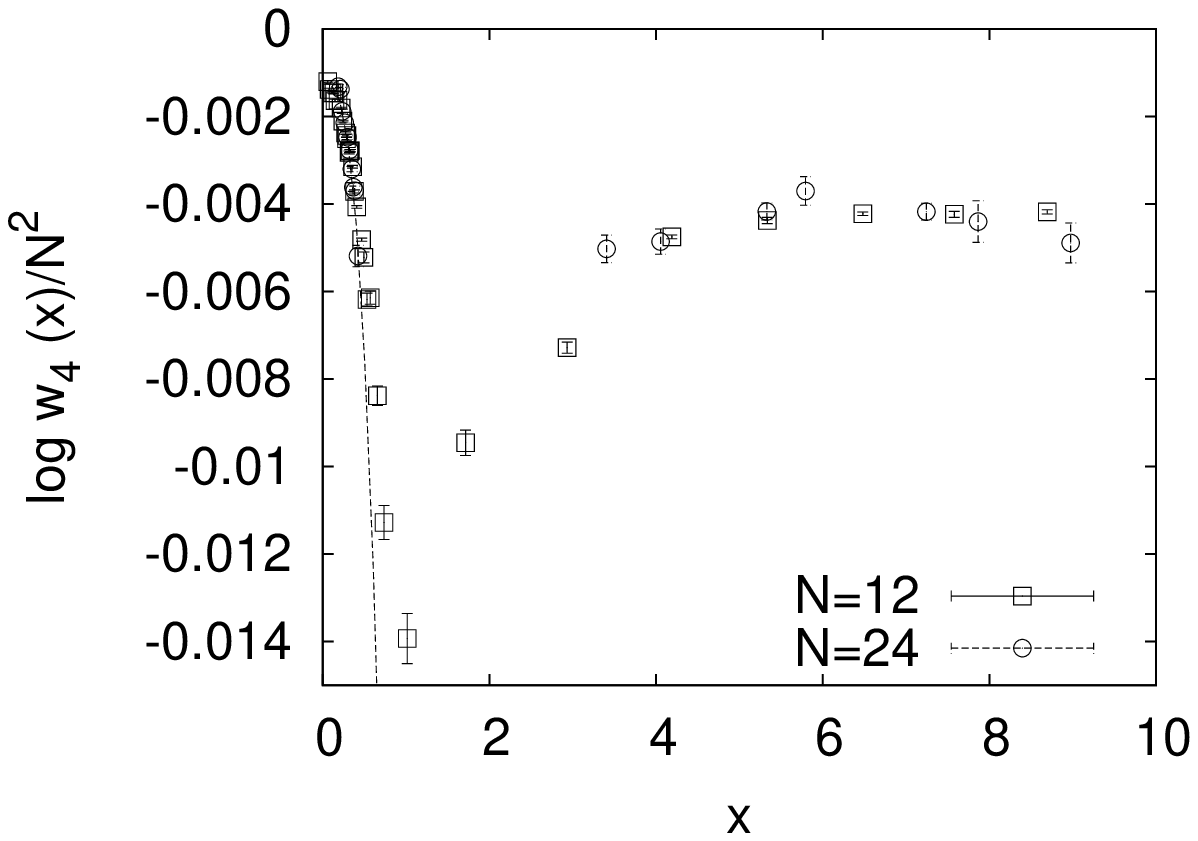}
\includegraphics[width=7.4cm]{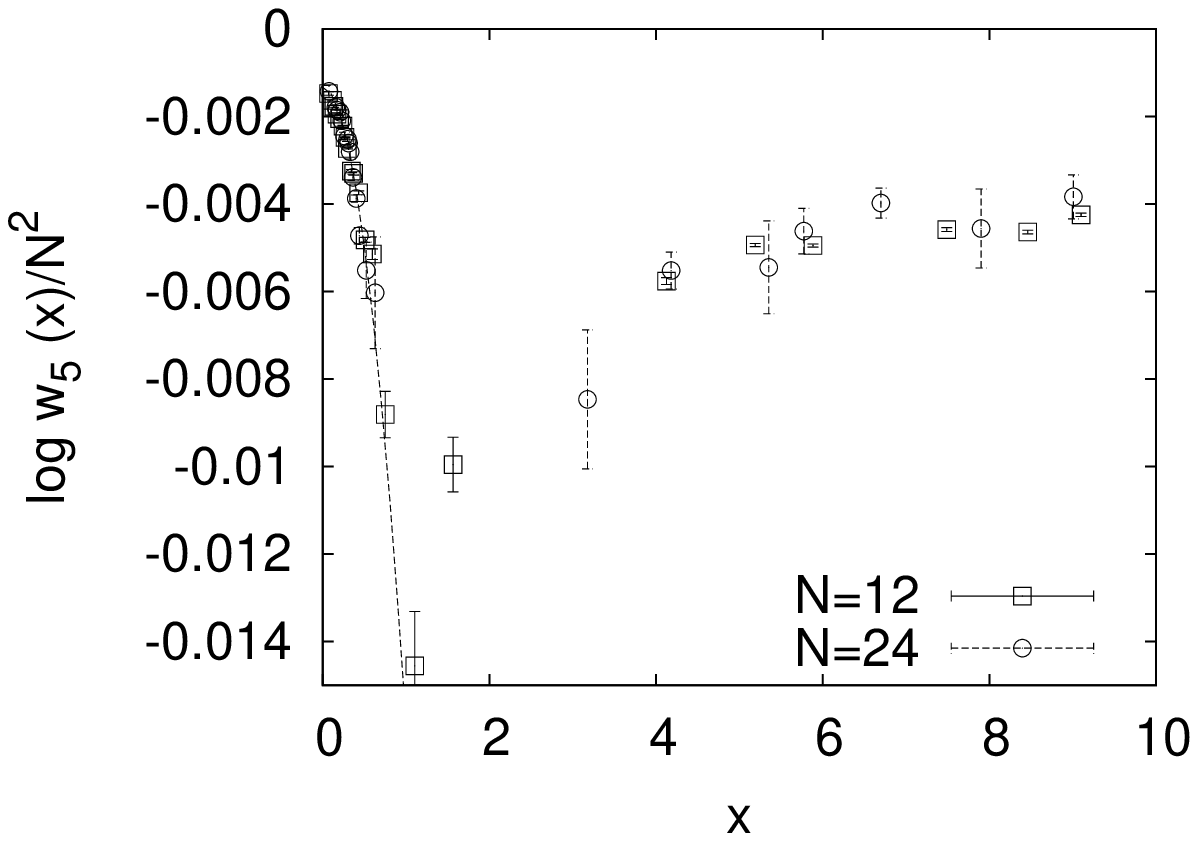}
\includegraphics[width=7.4cm]{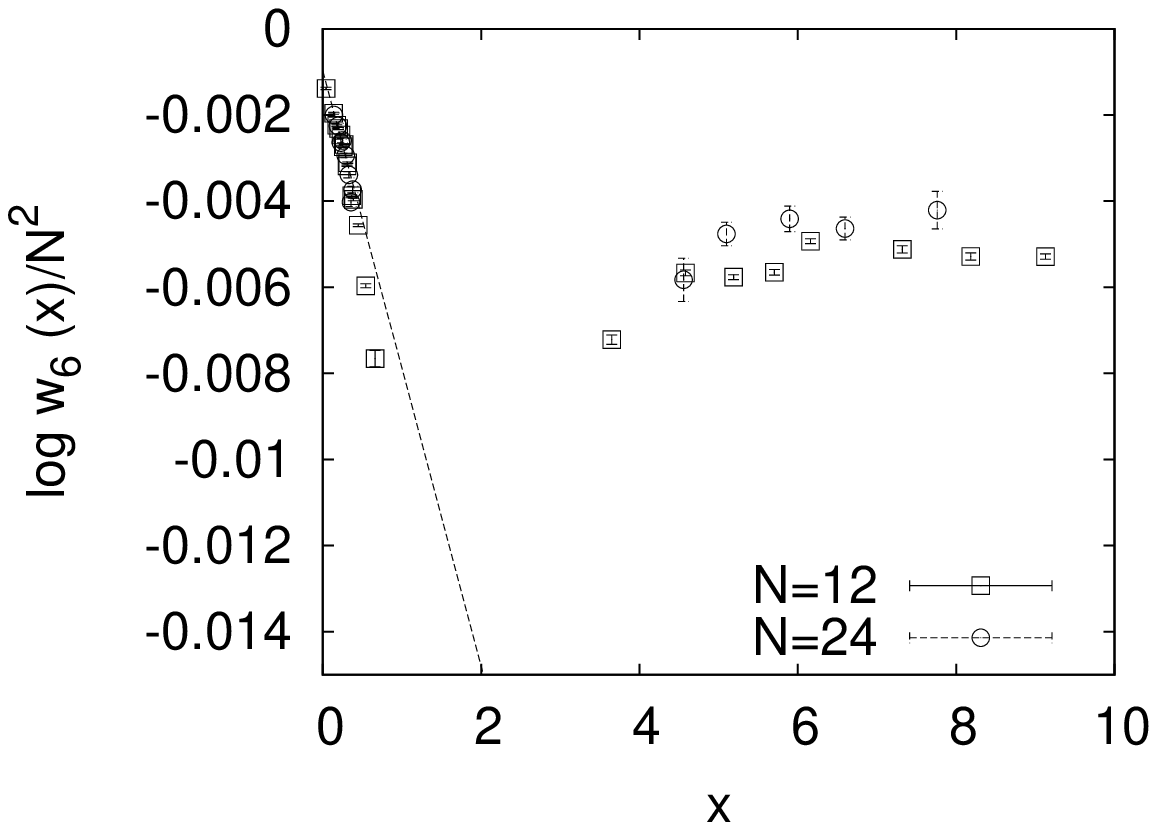}
    \caption{\label{f:02}
The function $\frac{1}{N^2} \log w_{n} (x)$ is plotted against $x$
      for $N=12, 24$ with $n=3,4,5,6$. The dashed lines are 
the fits to the scaling ansatz \protect\rf{3.9} 
obtained in figure \ref{f:02b}.
} 
\end{figure}

\begin{figure}[tbp]
\centering 

\includegraphics[width=7.4cm]{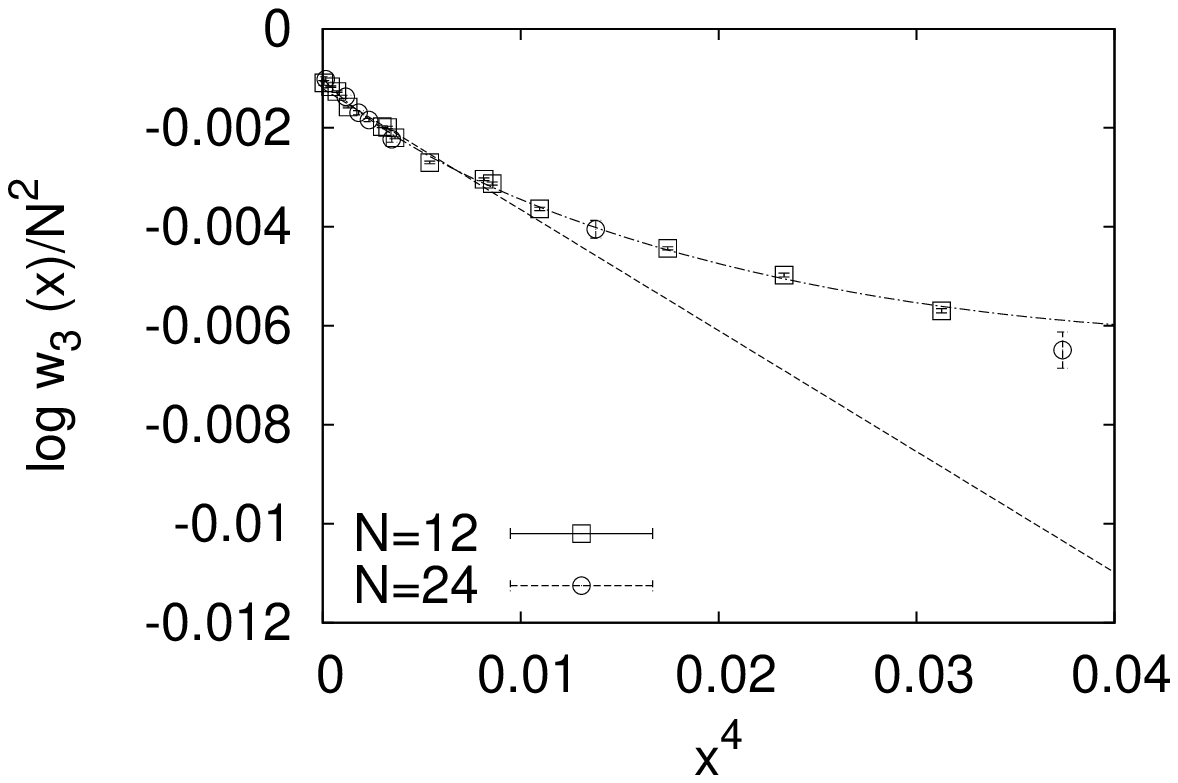}
\includegraphics[width=7.4cm]{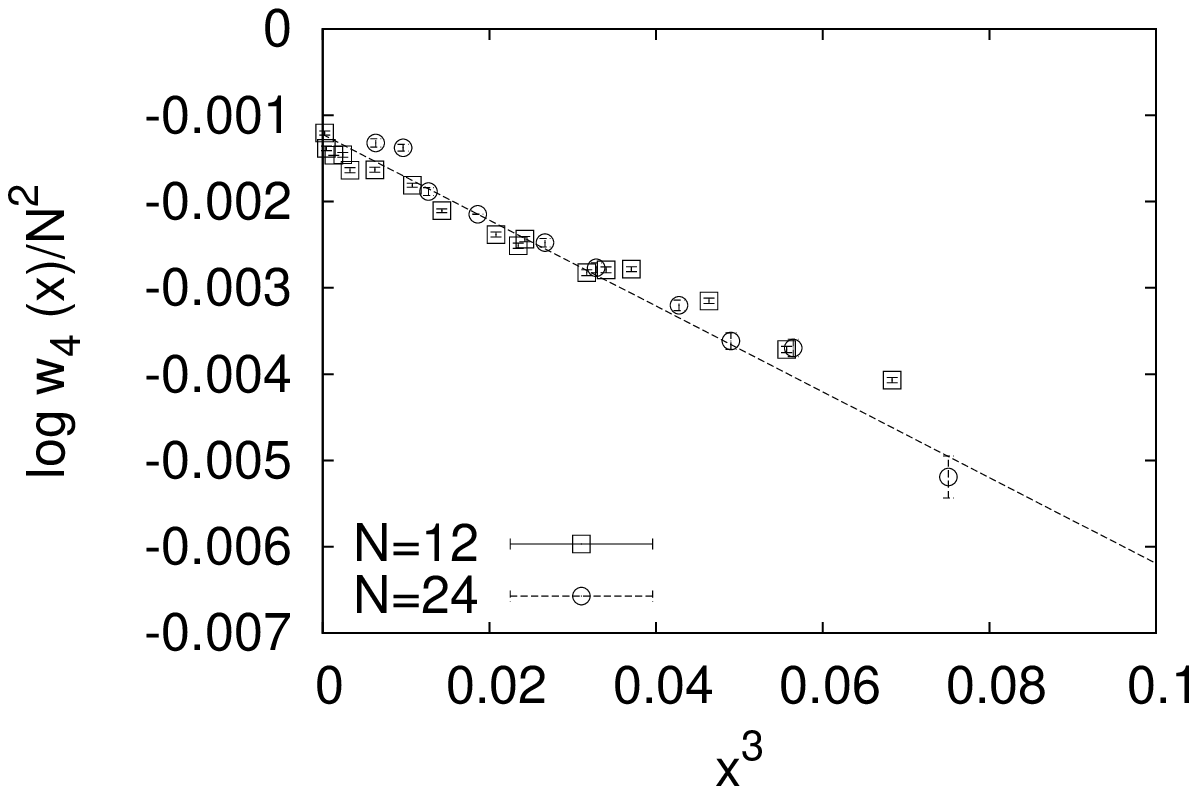}
\includegraphics[width=7.4cm]{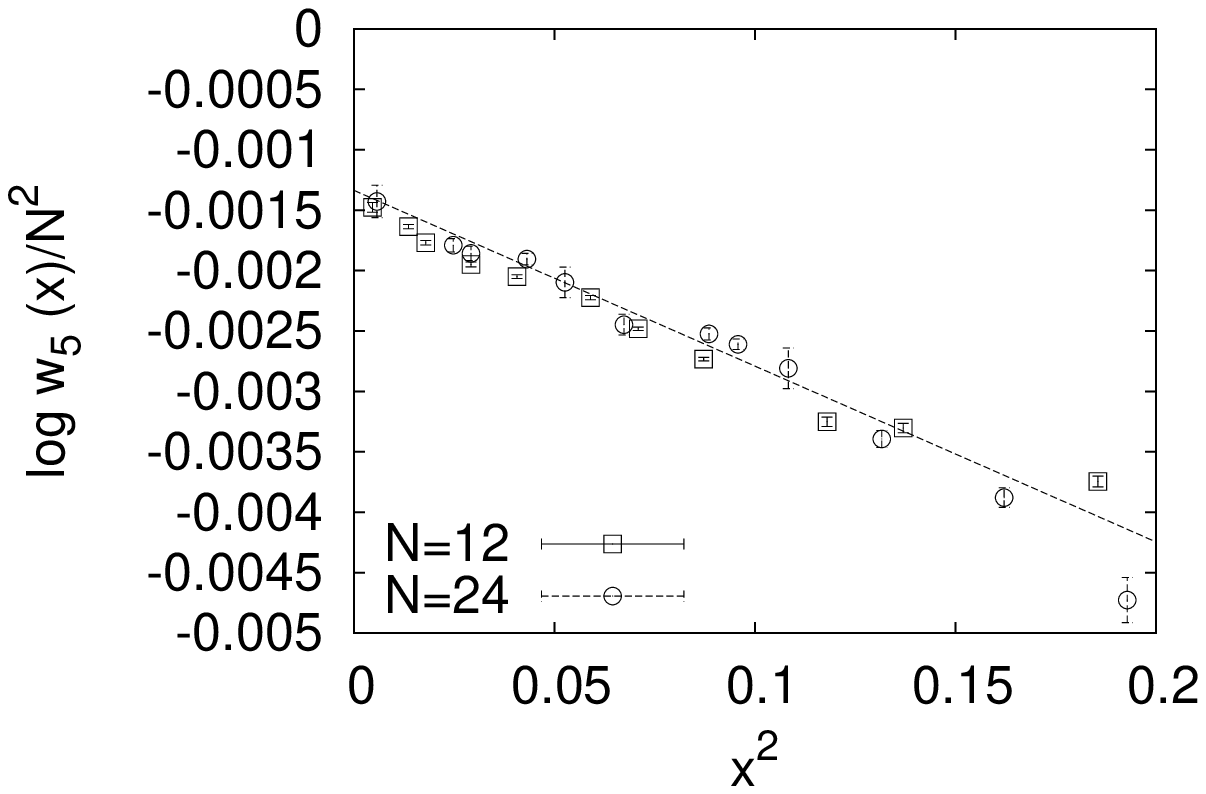}
\includegraphics[width=7.4cm]{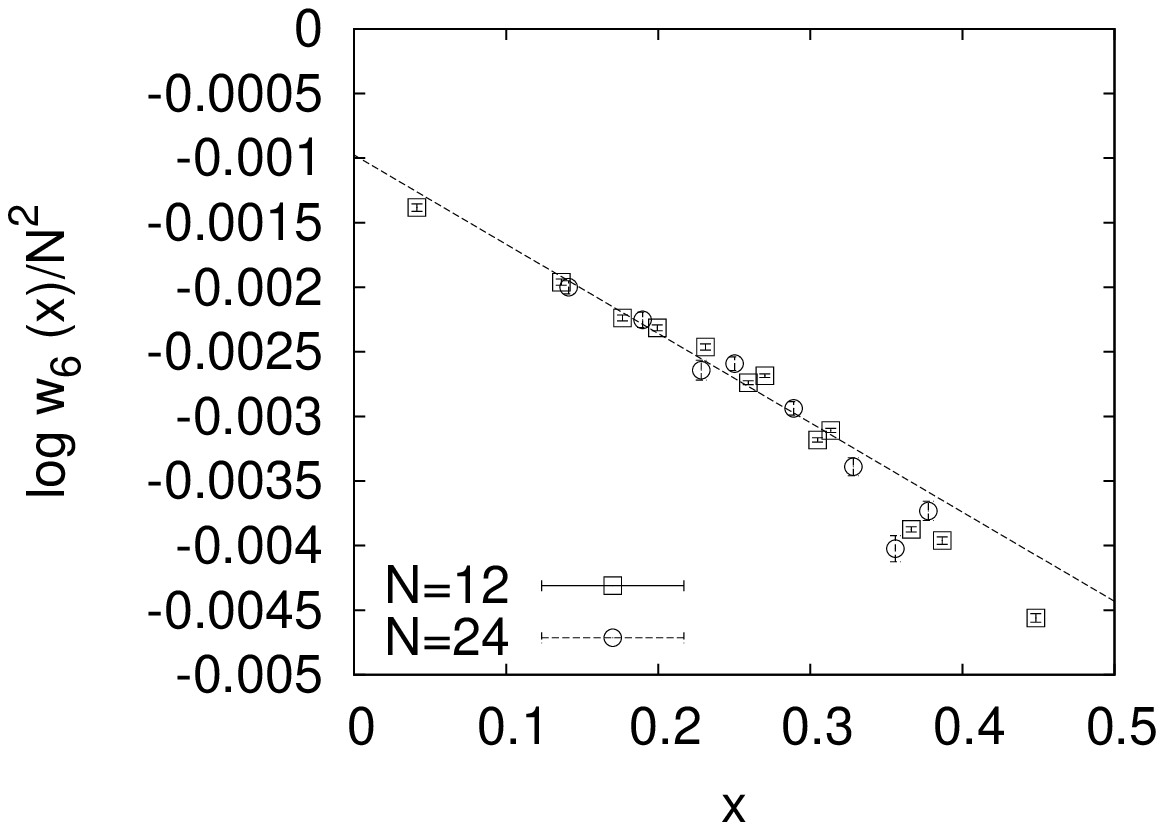}
    \caption{\label{f:02b}
The function $\frac{1}{N^2} \log w_{n} (x)$ is plotted against $x^{7-n}$
      for $N=12, 24$ with $n=3,4,5,6$. The dashed lines represent 
the fits of the $N=24$ data 
to the scaling ansatz \protect\rf{3.9}, which are used to compute
    $\Phi_n(x)$ in eq.~\protect\rf{3.7}.
The dash-dotted line in the top-left panel represents a fit
including a subleading term; 
See footnote \protect\ref{footnote:subleading}.
}
\end{figure}

Computing $w_n(x)$ using eq.~\rf{3.5} is hard due to the complex-action
problem,
especially near $x\approx 1$. The computation is easier in the region
near the solution to eq.~(\ref{2.30}),
where the fluctuations of the phase are milder. In such a region, one
can make a sensible large-$N$ extrapolation of
\begin{equation}
\label{3.7}
\Phi_n(x) = \lim_{N\to\infty} \frac{1}{N^2} \log w_n(x)\  .
\end{equation}
The above scaling is observed in figure \ref{f:02}, where we show our
results for $\frac{1}{N^2}\log w_n(x)$ with $n=3,4,5,6$ and $N=12,
24$.  The quality of convergence is similar to the oneloop model
studied in ref.~\cite{0108041} and the model studied in
refs.~\cite{Anagnostopoulos:2010ux,Anagnostopoulos:2011cn} as we have
explicitly checked for  $N=4, 6, 8, 12$ and $24$. Finite-$N$ effects
depend on the value of $x$ and for the purpose of
our calculation, it turns out that these need to be negligible in the
region $0.05<x<0.4$. By looking at figure \ref{f:02b} we see that this
has already been achieved with our data for $N=12, 24$, where finite-$N$
effects turn out to be smaller than other sources of systematic
errors. Then one can use the equation
\begin{equation}
\label{3.8}
\frac{1}{N^2} \, f^{(0)}_n(x) = - \frac{d}{dx}\Phi_n(x)
\end{equation}
in order to obtain the solution
$\bar{x}_n (< 1)$ to \rf{2.30} in the large-$N$ limit.
%
The left-hand side
of the above equation can be calculated without the 
complex-action
problem, and therefore we can obtain results at values
of $N$
larger than the ones used in order to determine
$\Phi_n(x)$. 
This feature of the factorization method turns out to be
crucial in the present work since 
$\frac{1}{N^2} \, f^{(0)}_n(x)$ suffers from
severe finite-$N$ effects 
in the interesting region of $x$ as we will see.
%
%

\begin{table}[tbp]
\centering
\begin{tabular}{|c|c|c|c|c|c|c|}
\hline
$n$ & $a_{n}\times 10^{2}$ 
             & $b_{n}\times 10^{3}$          & $c_n$   & $d_n$ 
  & $p_n$   & $q_n$         \\
\hline
 3  & $24(2) $           & $1.21(7)$  
    & $-2.7(3) $           & $2.2(7)$  
    &  $2.2(1) $    &  $14.0(3) $  \\
 4  & $5.0(5) $           & $1.2(1)$   
    & $-3.4(6) $           & $0(1)$   
    &  $4.6(2)$    &  $15.5(2)$ \\
 5  & $1.45(9)$           & $1.34(7)$  
    & $-3.7(3) $           & $-1.3(7)$   
    &  $4.3(3) $   &  $17.9(4) $  \\
 6  & $0.69(7) $           & $1.0(2) $    
    & $-3.6(3) $           & $-1.9(8)$   
    &  $3.0(3)$ &  $16.4(7)$ \\
\hline
\end{tabular}
\caption{\label{t:01}
The results for the fitting parameters are shown.
The parameters $a_{n}$ and $b_{n}$ in \protect\rf{3.9} 
are obtained from figure \ref{f:02b}.
The parameters
$c_{n}$ and $d_{n}$ in (\protect\ref{3.12})
are obtained from figure \ref{f:03}.
The parameters $p_{n}$ and $q_{n}$
in (\protect\ref{3.13}) are obtained
from figure \ref{f:04}.
}
\end{table}

The function $\Phi_n(x)$ has an asymptotic behavior in the $x\ll 1$ region
that can be easily understood geometrically \cite{Anagnostopoulos:2011cn}.
%
Note first that the dominant configurations in the ensemble (\ref{2.27})
for $x\ll 1$ have
$(7-n)$ shrunken directions and they are approximately $d=(n-1)$
dimensional. Since the phase of the determinant vanishes for collapsed
configurations (see discussion that leads to
eq.~\rf{2.8} and the footnote \ref{footnote:dim-five}),
$w_n(x) \equiv \vev{{\rm e}^{i\Gamma}}_{n,x}$ 
is expected naively to
approach $1$ as $x \rightarrow 0$.

From figures \ref{f:02} and \ref{f:02b}, we find that
$w_n(x)$ actually approaches a value
slightly smaller than $1$ as $x \rightarrow 0$.
This can be understood intuitively as follows.
As $x$ becomes small, $\lambda_k$ with $k=n,\ldots , 6$ become small,
but $\lambda_k$ with $k=1 , \ldots , (n-1)$ become large for
entropic reasons.
Then, due to the bosonic action $S_{\rm b}$ in eq.~\rf{2.2},
the matrices $A_\mu$ tend to become simultaneously diagonal 
up to SU($N$) symmetry.
For such configurations, the fermionic action $S_{\rm f}$ 
has quasi-zero-modes
corresponding to diagonal fermionic matrices, and the phase fluctuation
of the fermion determinant is enhanced.\footnote{For $n=3$, we also have
quasi-zero-modes associated with collapsed configurations
with $d \le 2$, which increases the phase fluctuations
as we mention below eq.~\rf{2.8}.
}
It is conceivable that this effect balances against
the suppression of the phase fluctuation due to collapsing space-time,
and leads to 
the small deviation from the expected asymptotic behavior.\footnote{The
small
deviation from the geometric argument does not
exist in the toy model studied 
in ref.~\cite{Anagnostopoulos:2010ux,Anagnostopoulos:2011cn},
which has a Gaussian action for the bosonic matrices instead of
a commutator squared term \rf{2.2}.
This is consistent with our intuitive explanation given here.
} 

In fact, the asymptotic behavior of $w_n(x)$ for $x\ll 1$ 
can be fitted well to
\begin{equation}
\label{3.9}
\frac{1}{N^2} \log w_n(x) \simeq - a_{n}\, x^{7-n} - b_{n}
\quad \mbox{for} \quad
x\ll 1 \  , \quad n=2,3,4,5,6
\end{equation}
as shown in figure \ref{f:02b}.
The coefficients $a_{n}$ and $b_{n}$ obtained by the fits are
given
in table \ref{t:01}. 
The existence of a small positive constant $b_{n}$ 
can be attributed to the
effect just mentioned.
The first power-law term in eq.~(\ref{3.9})
can be derived \cite{Anagnostopoulos:2011cn}
from eq.~\rf{2.8}, which implies that the fluctuation of
the phase around a collapsed configuration with $1\le d\le 5$ is of
the order of $\delta\Gamma \sim (\delta A/|A|)^{6-d}$, where $\delta
A$ and $|A|$ represent a typical scale of $A_\mu$ in the shrunken and
extended directions, respectively. From eq.~\rf{2.9} we expect that
$\delta A/|A| \propto \sqrt{x}$ and that the width of the distribution
of the phase is $\sigma\propto (\sqrt{x})^{7-n}$. Assuming that the
distribution is Gaussian, the 
VEV of ${\rm e}^{i\Gamma}$
is given by
$\displaystyle
\int\, d\Gamma\,\frac{1}{\sqrt{2\pi}\sigma}
 \exp{\left(-\frac{1}{2\sigma^2}\Gamma^2\right)}
 {\rm e}^{i \Gamma} = \exp{\left(-\frac{1}{2}\sigma^2\right)}\  ,
$
which gives $-\log w_n(x)=\dfrac{1}{2}\sigma^2\propto x^{7-n}$. 

Let us also comment on the large-$x$ behavior of 
$ w_n (x)$ 
although it is not
of our primary interest in the present analysis
since we are searching for solutions to
eq.~(\ref{2.30}) in the $x<1$ region.
From figure \ref{f:02}, we find that
$ w_n (x)$ 
increases and 
approaches a constant at sufficiently large $x$.
This can be understood as follows.
In the large-$x$ regime, $\lambda_{k}$ with $k=1,\ldots , n$
are forced to be large,
and due to the bosonic action $S_{\rm b}$ in eq.~\rf{2.2},
the matrices $A_\mu$ tend to become simultaneously diagonal 
up to SU($N$) symmetry.
Then, neglecting the quasi-zero-modes discussed above eq.~\rf{3.9},
the fermion determinant becomes positive definite\footnote{For 
$n \le 5$, the dominant configuration
in the large-$x$ regime is $n$-dimensional, which adds to the
suppression of the phase fluctuation. This effect is expected
to become stronger for smaller $n$, which is seen in the
behavior of $\frac{1}{N^2} \log w_n (x)$ 
as $x$ decreases towards $x=1$
in figure \ref{f:02}.}
\begin{equation}
\det{\cal M}[A]= \left[\prod_{i < j} 
(\alpha_{i \mu} - \alpha_{j \mu})^2
\right]^4  > 0 \ ,
\end{equation}
where $A_\mu = {\rm diag}(\alpha_{1\mu} , \ldots ,
\alpha_{N \mu})$.
Due to the existence of the quasi-zero-modes, however,
the VEV of ${\rm e}^{i\Gamma}$ becomes slightly below 1
for the same reason as before.\footnote{Figure \ref{f:02}
shows that the constant value to which $w_n(x)$ approaches 
at large $x$ depends very little on $n$.
This is understandable since the value is determined
by the dynamics of the quasi-zero modes represented
by diagonal fermionic matrices, where the dimensionality
of configurations does not play important roles. 
}
As a result,
the dependence of 
$ w_n (x)$ 
on $x$ becomes quite
mild in the large-$x$ region.
This leads to our assumption in section \ref{sec:method}
that the large eigenvalues $\lambda_n$ 
do not affect much the fluctuations of the phase.

Let us move on to the computation of the left-hand side of eq.~\rf{3.8}.
First we note that the small-$x$ behavior
of the function $\dfrac{1}{N^2}
f^{(0)}_n(x)$ is expected to be\footnote{In the $n=1$ case, 
which we do not study here, 
we have to consider that the eigenvalues of $A_\mu$ 
collapse to $0$ for $x\ll 1$,
and the suppression factor comes also from the fermion determinant,
which is a homogeneous polynomial of $A_\mu$ of degree $4 (N^2-1)$. This
gives an extra suppression factor of $(\sqrt{x})^{4 (N^2-1)}$.
Hence eq.~(\ref{3.11}) should be replaced by
$\lim _{N\rightarrow \infty} 
\frac{1}{N^2} f^{(0)}_1(x) \simeq  \frac{5}{x}$.
}
\begin{equation}
\label{3.11}
\lim _{N\rightarrow \infty}
\frac{1}{N^2} f^{(0)}_n(x) \simeq
  \frac{7-n}{2x}
\  .
\end{equation}
This can be understood
\cite{Anagnostopoulos:2010ux,Anagnostopoulos:2011cn} in terms of the
``phase-space suppression'' since 
$(7-n)$ directions shrink as $x$ becomes small. 
Each direction has an extent $\sim \sqrt{x}$, which results in 
$\rho^{(0)}_n(x)\sim (\sqrt{x})^{(7-n)(N^2-1)}$. 
From the definition \rf{2.30}
of $f^{(0)}_n(x)$,
we obtain eq.~\rf{3.11}.

\begin{figure}[tbp]
\centering 
\includegraphics[width=7.4cm]{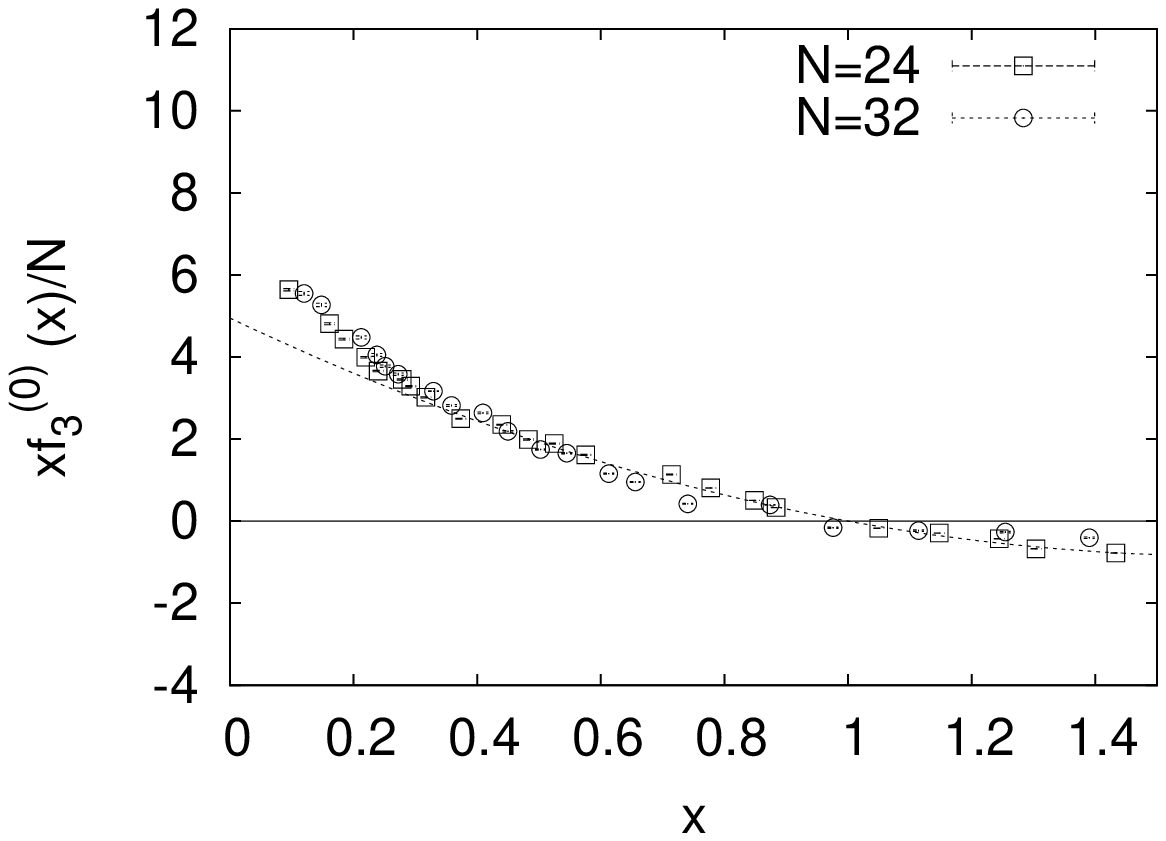}
\includegraphics[width=7.4cm]{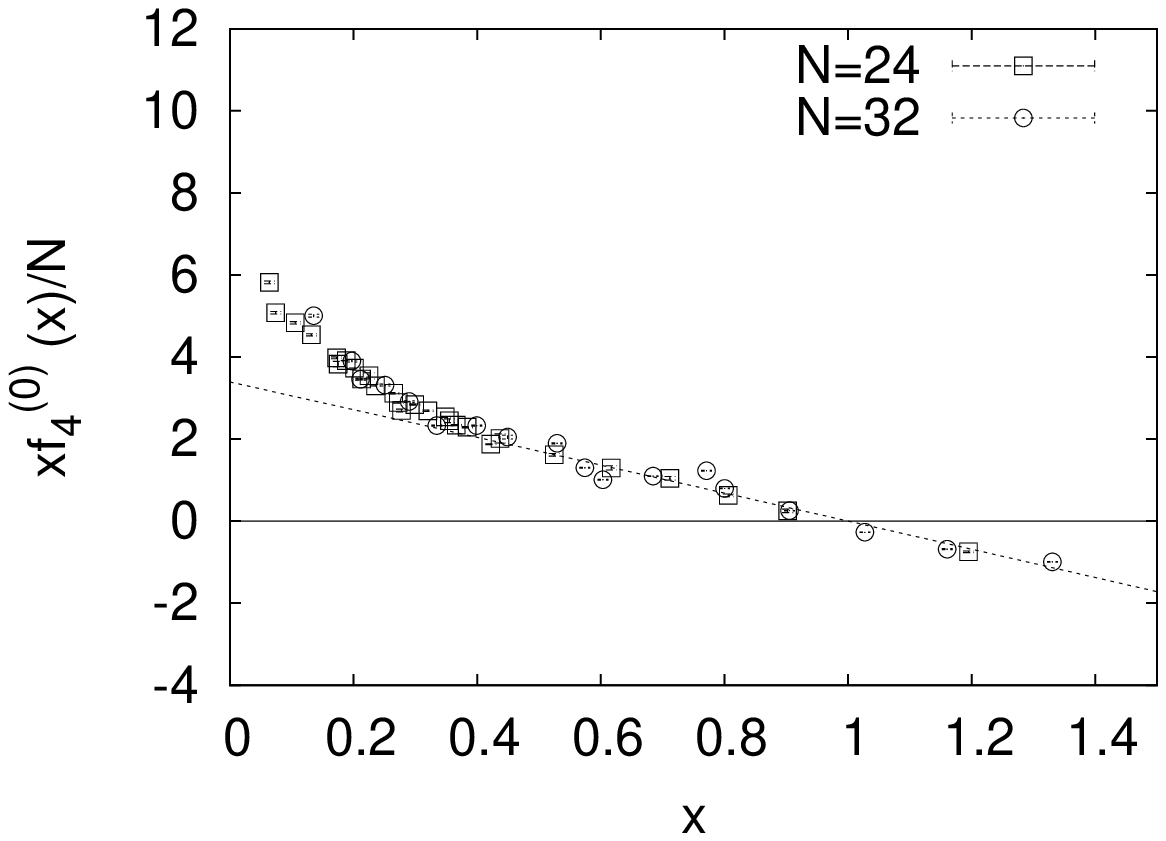}
\includegraphics[width=7.4cm]{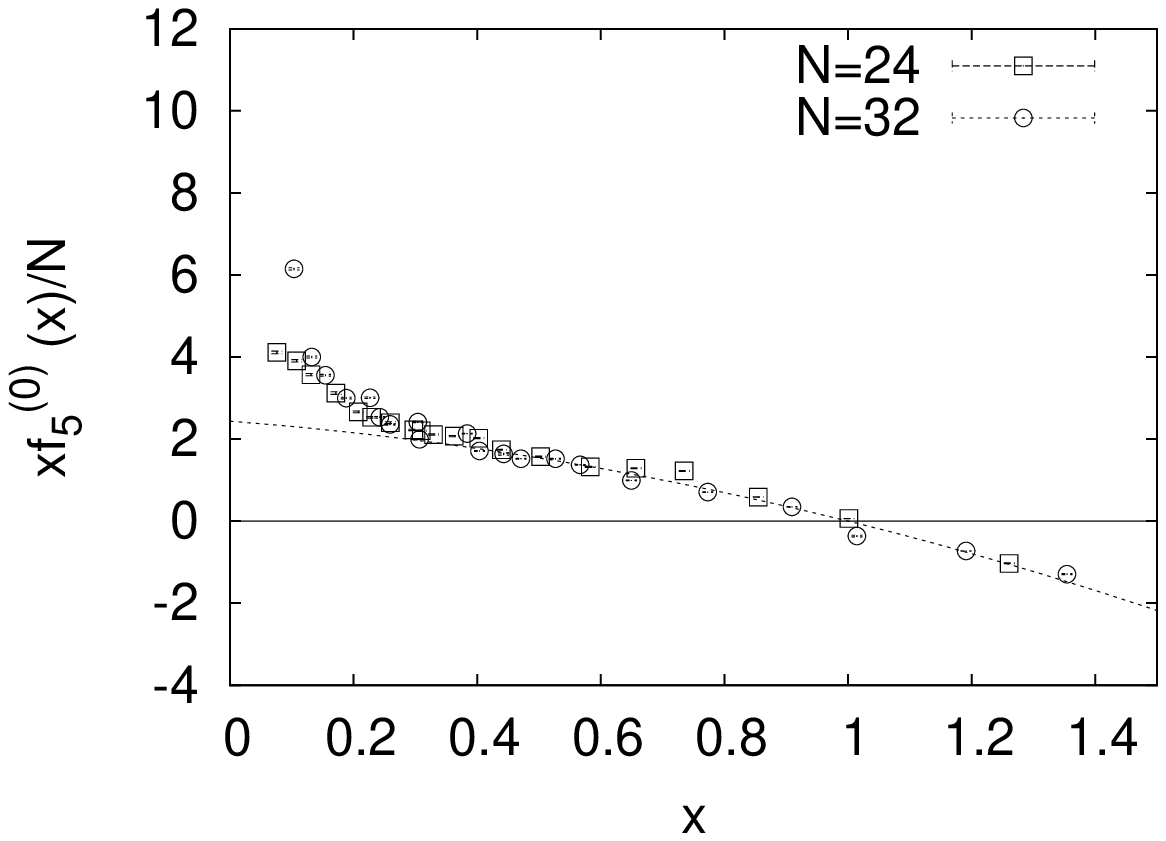}
\includegraphics[width=7.4cm]{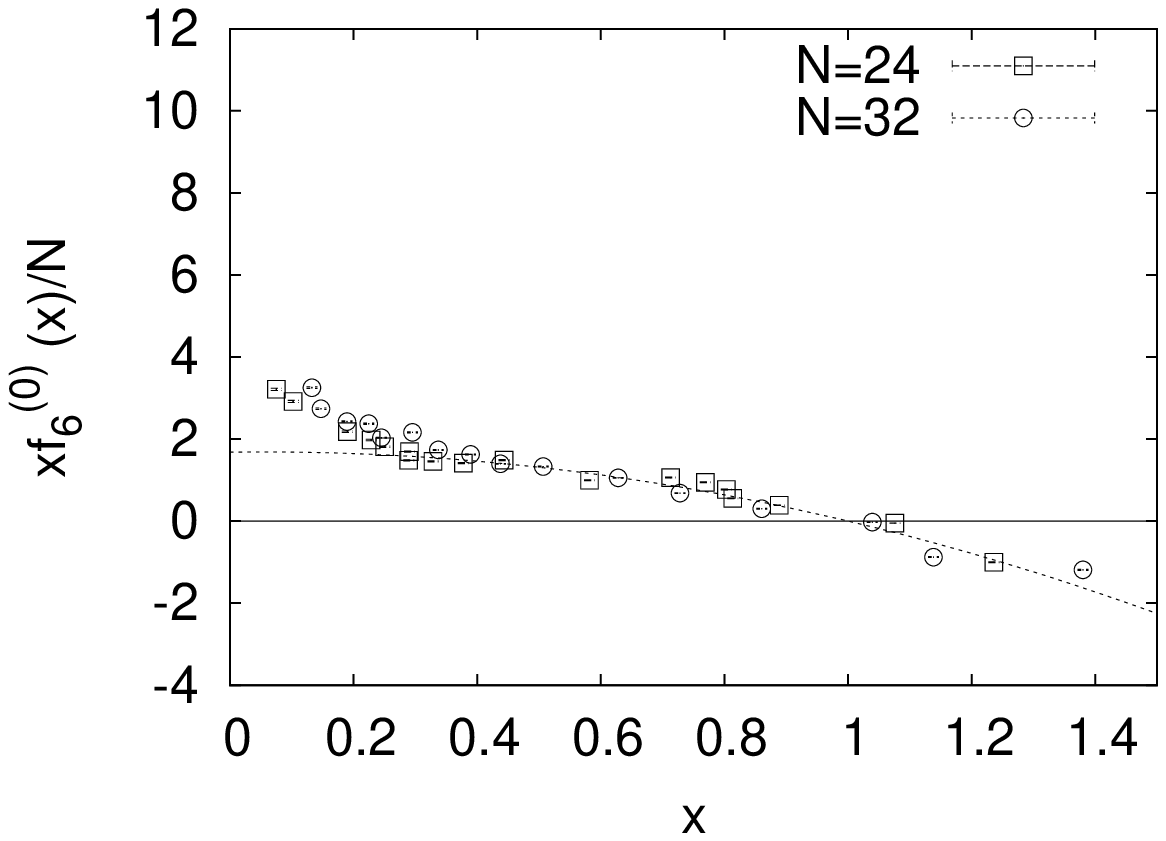}
    \caption{\label{f:03}
The function $\frac{x}{N} f^{(0)}_{n} (x)$ is plotted against
$x$ for $N=24, 32$ with $n=3,4,5,6$. 
The dotted lines represent the fits of all the data within $0.4 \le x \le 1$
to the function $g_n(x)$ in eq.~(\protect\ref{3.12}).
}
\end{figure}

In the region $x\gtrsim 0.4$, however, we find a completely
different large-$N$ scaling as shown in figure \ref{f:03}.
It is actually $\dfrac{1}{N}f^{(0)}_n(x)$ that scales,
which implies that $\dfrac{1}{N^2}f^{(0)}_n(x)$ vanishes 
as ${\cal O}(\frac{1}{N})$ in this region.\footnote{This reduction 
in the values of $\dfrac{1}{N^2}f^{(0)}_n(x)$ is not observed
in the matrix model studied in
ref.~\cite{Anagnostopoulos:2010ux,Anagnostopoulos:2011cn},
which has no supersymmetry.}
This can be understood
as a result of cancellations by fermionic and bosonic contributions to
the interactions among space-time points\footnote{The 
remaining ${\cal O}(\frac{1}{N})$ terms in $\dfrac{1}{N^2}f^{(0)}_n(x)$ 
can be attributed to the branched-polymer-like interaction \cite{9802085}
among the space-time points, which arises from the integration
over the fermionic zero modes with only ${\cal O}(N)$ degrees of
freedom. See also footnote \ref{footnote:flat-direction}.}
as was first noted in the oneloop approximated model studied in
ref.~\cite{0108041}. 
In that model, however, the $\dfrac{1}{N}f^{(0)}_n(x)$
scaling was observed also in the small-$x$ region.
This shows that the non-vanishing $\dfrac{1}{N^2}f^{(0)}_n(x)$ 
with the small-$x$ behavior (\ref{3.11})
arises in the full model considered here
due to the small distance interactions
of the eigenvalues, which are ignored in the oneloop model.

\begin{figure}[tbp]
\centering 
\includegraphics[width=7.4cm]{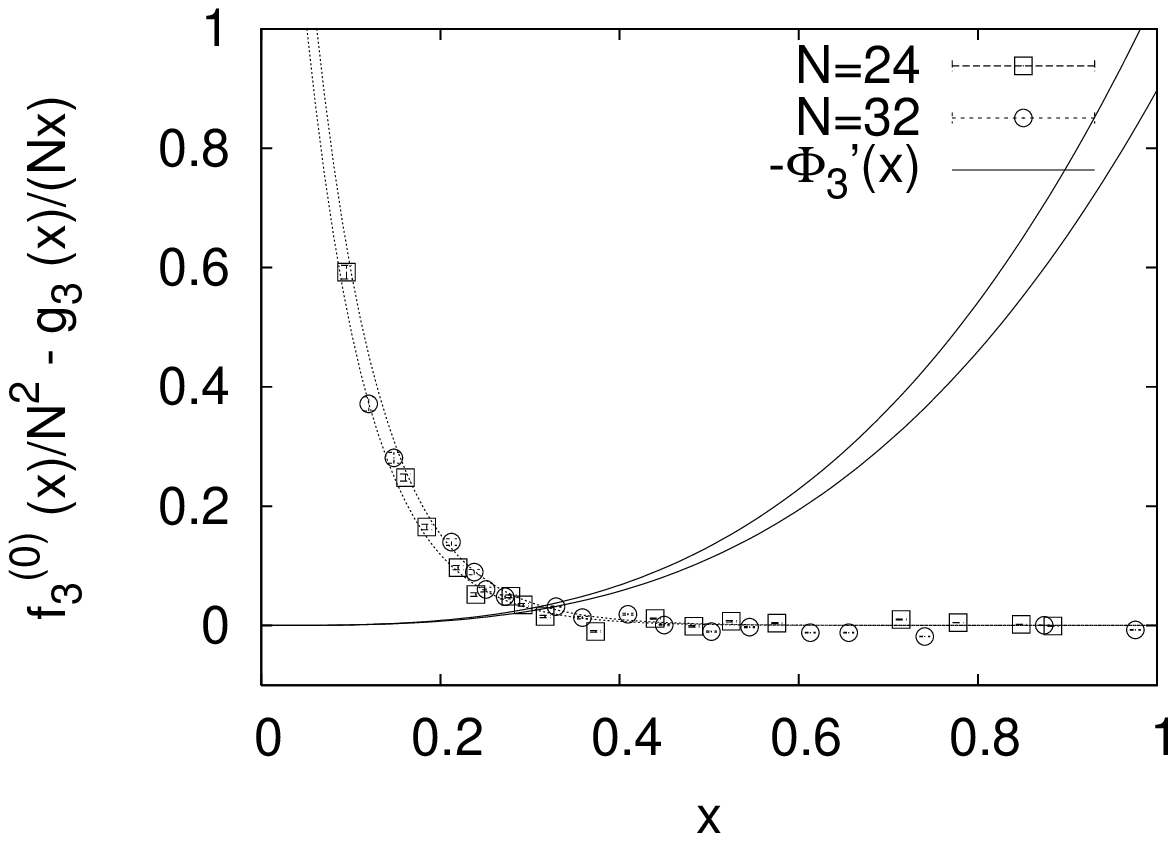}
\includegraphics[width=7.4cm]{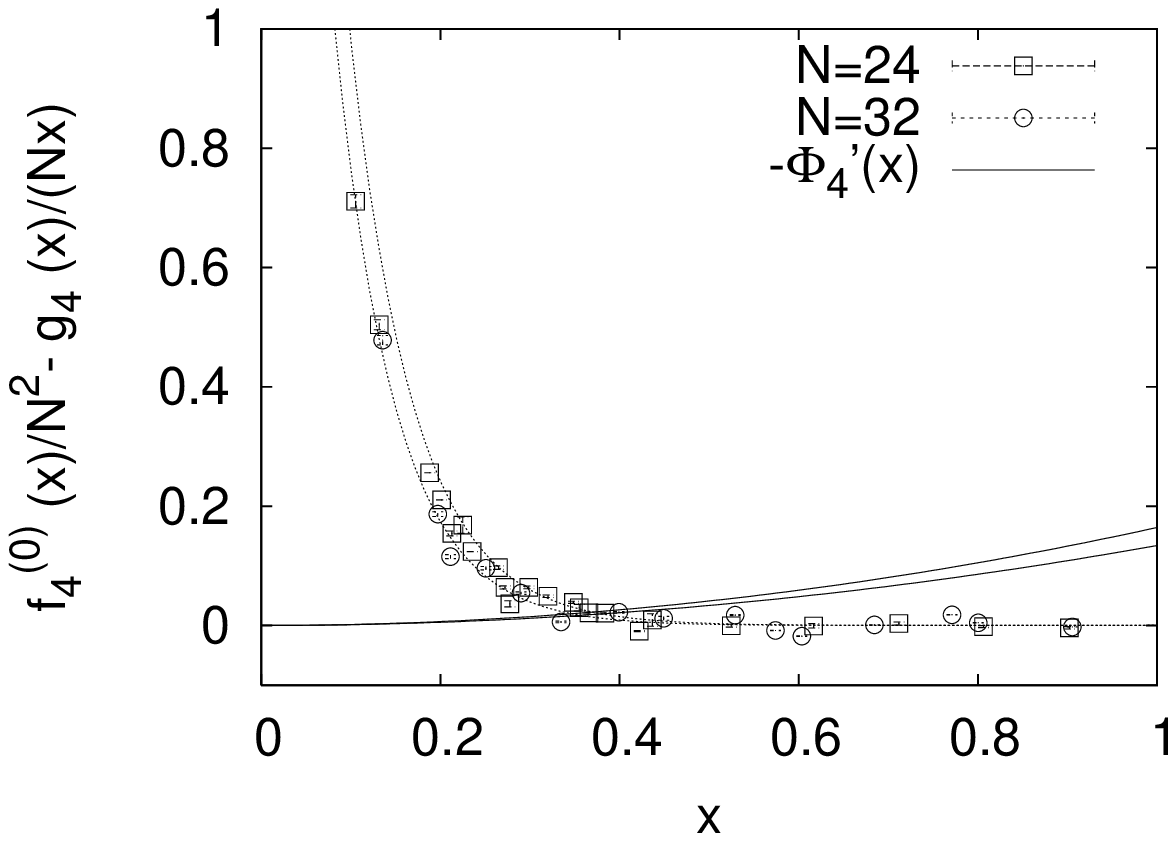}
\includegraphics[width=7.4cm]{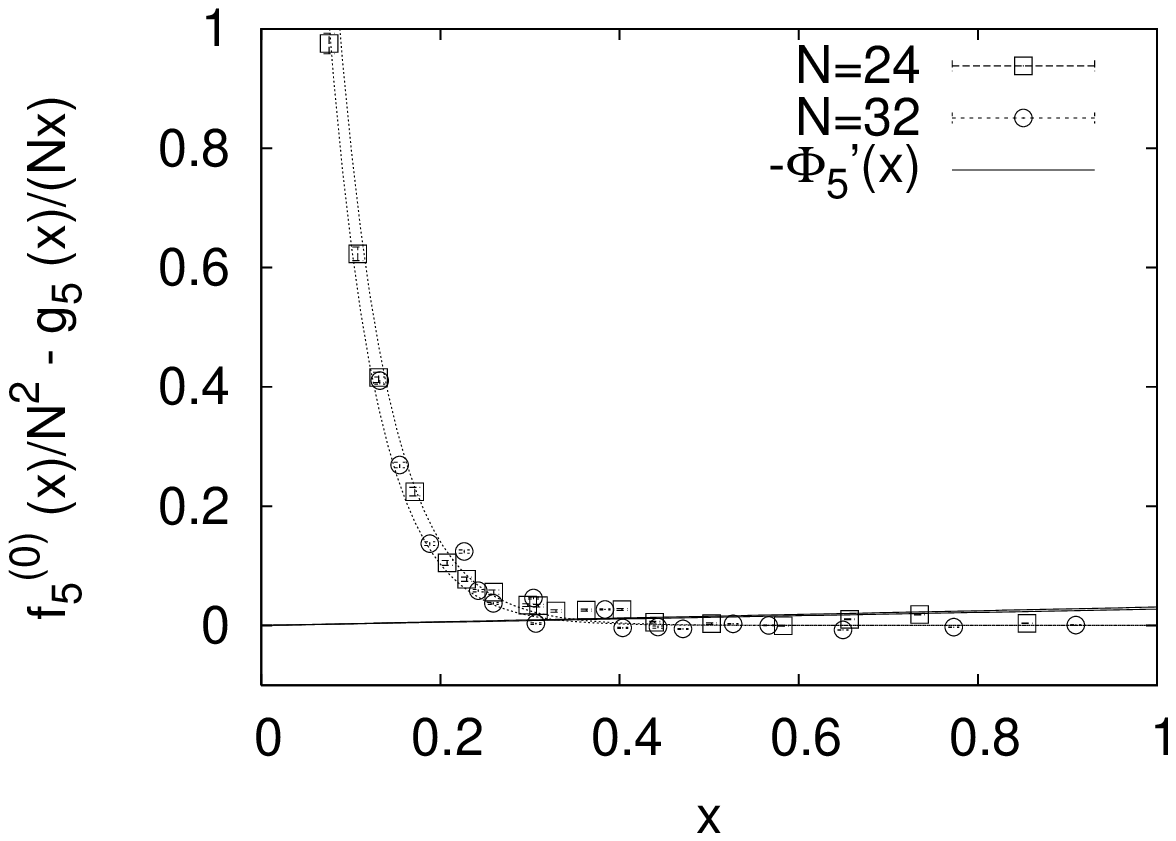}
\includegraphics[width=7.4cm]{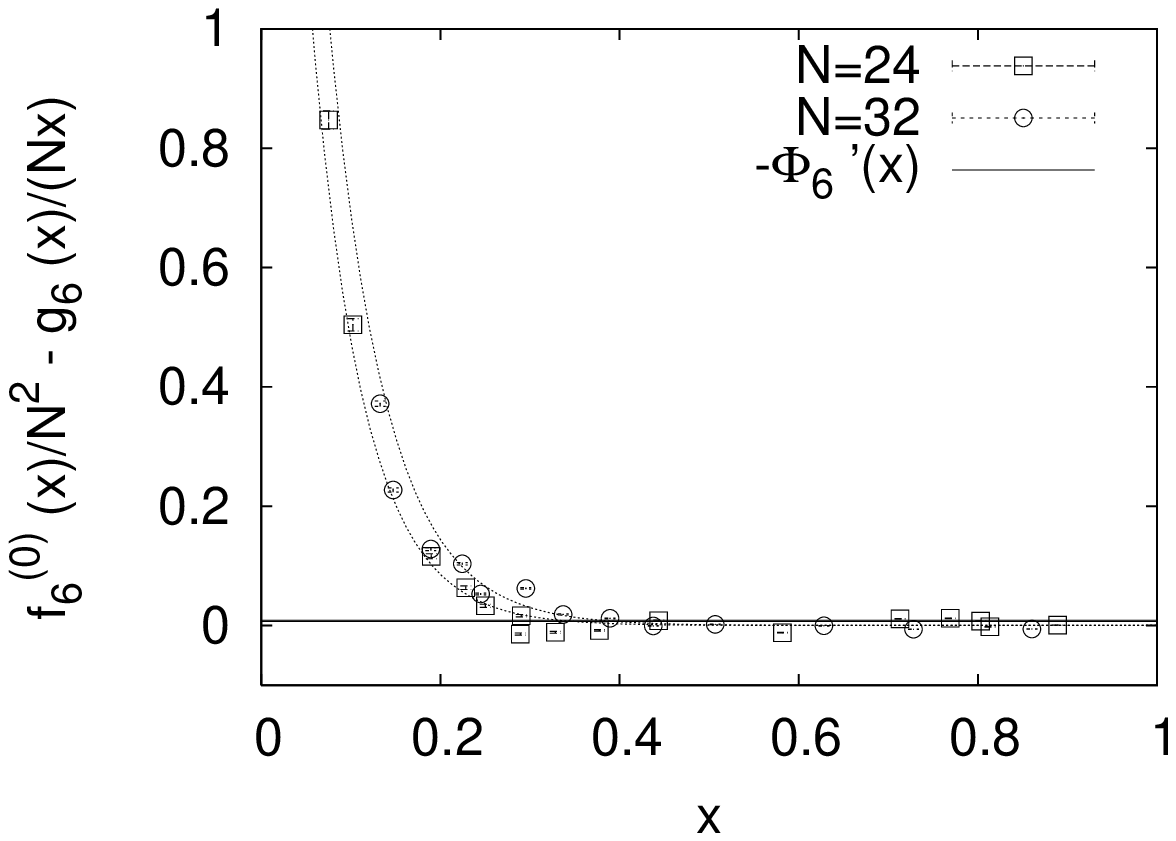}
    \caption{\label{f:04}
The function $\frac{1}{N^2} f^{(0)}_{n} (x) - \frac{g_n(x)}{Nx}$
is plotted against $x$ for $N=24, 32$ with $n=3,4,5,6$.
The dotted lines represent
the fits of all the data within $0.1 \le x \le 0.4$
to the ansatz (\protect\ref{3.13}).
The solution to
eq.~\protect\rf{3.8} is then given by the intersection with the
function $-\Phi_{n}'(x)$ represented by the solid lines with a margin.}
\end{figure}


The existence of the ${\cal O}(\frac{1}{N})$ terms in
$\dfrac{1}{N^2}f^{(0)}_n(x)$ obscures the leading large-$N$ behavior,
and a direct large-$N$ extrapolation would require much larger $N$
than we can study.
Therefore we attempt to subtract the ${\cal O}(\frac{1}{N})$ terms in the
following way.
From figure \ref{f:03} we find that the data for 
$\dfrac{1}{N}f^{(0)}_n(x)$ within $0.4 \le x \le 1$
can be fitted nicely to
\begin{equation}
\label{3.12}
\frac{x}{N} f^{(0)}_n(x) \simeq g_n(x) =
 c_{n}(x-1)+ d_{n}(x-1)^2
\  ,
\end{equation}
which implies a near Gaussian behavior of $\rho^{(0)}_n(x)$ around $x=1$.
The coefficients $c_n$ and $d_n$
obtained by the fits are given in table \ref{t:01}.
We subtract the ${\cal O}(\frac{1}{N})$ terms in 
$\dfrac{1}{N^2}f^{(0)}_n(x)$ 
given by eq.~(\ref{3.12}) from our data,
and show the results in figure \ref{f:04}.
We find that the results for $N=24$ and $N=32$ scale reasonably well,
and the data within $0.1 \le x \le 0.4$
can be fitted to a simple ansatz
\begin{equation}
\label{3.13}
\frac{1}{N^2} f^{(0)}_n(x) - \frac{g_n(x)}{N x} =
p_n \exp{(-q_n x)} \  
\end{equation}
as shown in
figure \ref{f:04}. 
The parameters $p_n$ and $q_n$ obtained by the fits
are given in table \ref{t:01}.

In the same figure we also 
plot $-\Phi'_n(x)$, where we use eq.~\rf{3.9}
with $a_n$ and $b_n$ shown in table \ref{t:01}
as an estimate of $\Phi_n(x)$.
The values of $x$
at the intersections with the function \rf{3.13}
are given as\footnote{The value of $\bar{x}_3$
quoted 
here
is obtained by taking into
account the deviation of
$\frac{1}{N^2} \log w_3(x)$ from the asymptotic behavior (\ref{3.9})
at $x\gtrsim 0.3$ seen in the top-left panel of figure~\ref{f:02b}.
As an estimate of $\Phi_3(x)$, we have actually used
the dash-dotted line in the same panel, which
represents a fit including 
a subleading term $\propto x^{9/2}$ \cite{Anagnostopoulos:2011cn}. 
If we use instead the dashed line 
representing a fit to (\ref{3.9}),
we obtain
$\bar{x}_3 = 0.31(1)$.\label{footnote:subleading}}
\beq
\bar{x}_3 = 0.33(1) \ , \quad \bar{x}_4 = 0.35(1) \ , \quad
\bar{x}_5 = 0.34(2) \ , \quad \bar{x}_6 = 0.36(3) \ . 
\label{main-results}
\eeq
These values of $\bar{x}_n$ with $n=d+1$
provide estimates for 
$\vev{\tilde\lambda_{d+1}}_{{\rm SO}(d)}$ 
in the SO($d$) symmetric vacua as explained below eq.~(\ref{2.30}).
Recalling the adopted normalization (\ref{2.13})
and our observation 
$\lim_{N\rightarrow \infty} \vev{\lambda_n}_0 = \ell^2$
from figure \ref{f:01}, we find that 
our results for $\vev{\tilde\lambda_{d+1}}_{{\rm SO}(d)}$ 
are
in good agreement with
\beq
\vev{\tilde\lambda_{d+1}}_{{\rm SO}(d)} 
\equiv \frac{\vev{\lambda_{d+1}}_{{\rm SO}(d)}}
{\vev{\lambda_{d+1}}_{0}}
= \frac{r^2}{\ell^2} 
\approx \frac{0.223}{0.627} = 0.355 \ ,
\label{GEM-pred-ratio}
\eeq
where we used the GEM predictions \cite{10070883}
reviewed at the end of section \ref{sec:model}.
Thus our results are consistent with the 
universal scale of small dimensions for all the SO($d$) symmetric
vacua with $2 \le d \le 5$.

\begin{figure}[tbp]
\centering 
\includegraphics[width=9cm]{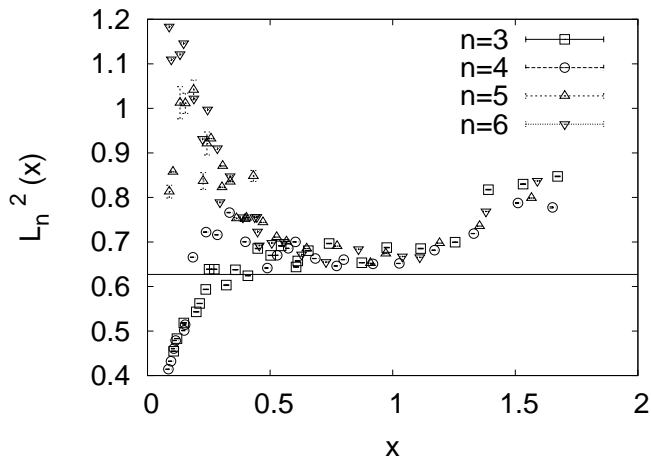}
    \caption{\label{f:05}The geometric mean 
$L_n ^2(x) = \left( \prod_{k=1}^{6} \langle
      \lambda_{k} \rangle_{n,x} \right)^{\frac{1}{6}}$ is plotted for
      $N=32$ with $n=3,4,5,6$. The solid line represents $\ell^2 =0.627$,
which appears 
in eq.~(\ref{1.1}) obtained by the GEM \cite{10070883}.
} 
\end{figure}

Let us turn to the calculation of the large eigenvalues.
The eigenvalues $\lambda_k$ in the SO($d$) symmetric vacua
can be estimated by eq.~\rf{vev-lambda}.
More generally, we consider $\vev{\lambda_k}_{n,x}$
for various $x$, and calculate them by
\begin{equation}
\vev{\lambda_k}_{n,x_{\rm p}} = 
\vev{\lambda_k}_{n,V} \ ,
\end{equation}
where the right-hand side can be
measured in the system \rf{3.1},
and $x_{\rm p}$ on the left-hand side is defined by eq.~\rf{3.4}.
In particular, we calculate the geometric mean
\begin{equation}
\label{3.15}
L_n^2 (x) = \left(
 \prod_{k=1}^6 \VEV{\lambda_k}_{n,x}
 \right)^{\frac{1}{6}} \ .
\end{equation}
The results are shown in figure \ref{f:05}, where we see
that $L_n^2(x)\approx \ell^2 \approx 0.627$ within $0.5 < x < 1$
for all $n$, which is consistent with the implication 
of the constant volume property (\ref{1.1})
discussed at the end of section \ref{sec:model}.
%
Let us note that, for $x=1$, 
the system essentially becomes the phase-quenched model
without any constraint, and therefore we have
$\vev{\lambda_k}_{n,x=1} \approx \vev{\lambda_k}_{0}$ for all $n$.
In view of the results in figure \ref{f:01},
the deviation of $L_n^2(x=1)$ from $\ell^2$ can be 
attributed to finite-$N$ effects.
Eq.~(\ref{1.1}) for the SO($d$) symmetric vacua 
with $3 \le d \le 5$ 
suggests that the constant volume property 
extends to smaller $x$ including $\bar{x}_n \approx 0.355$
for $4 \le n \le 6$.
This is not clearly seen in figure \ref{f:05}
presumably due to finite-$N$ effects, however.

\begin{figure}[tbp]
\centering 
\includegraphics[width=9cm]{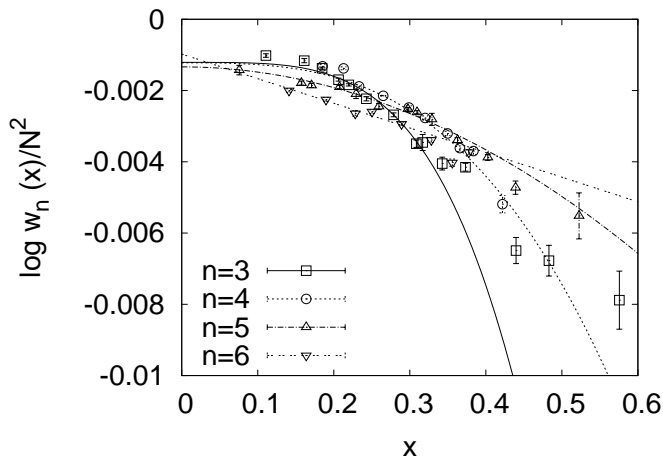}
    \caption{\label{f:06}
The function $\frac{1}{N^2} \log w_{n} (x)$ with $n=3,4,5,6$ are 
plotted together for $N=24$. 
} 
\end{figure}

Finally, we attempt to compare the free energies 
for the SO($d$) vacua ($d=2,3,4,5$) 
using eq.~(\ref{simpler-free-energy}).
For simplicity, let us assume that $\bar{x}_n = 0.355$ 
as suggested by the GEM.
The first term in eq.~(\ref{simpler-free-energy})
can be estimated as
\begin{equation}
\int _{\bar{x}_n}^1  \frac{1}{N^2}f^{(0)}_n(x)  dx
= \frac{p_n}{q_n} ({\rm e}^{-q_n \bar{x}_n } - {\rm e}^{-q_n }) \ ,
\label{first-termF}
\end{equation}
where we have used the right-hand side of eq.~\rf{3.13}
as an estimate for the large-$N$ limit of $\frac{1}{N^2}f^{(0)}_n(x)$.
Plugging in the values of 
$p_n$ and $q_n$ given in table \ref{t:01}, 
eq.~\rf{first-termF} is estimated as
0.001, 0.001, 0.0004, 0.0005
for $n=3,4,5,6$, respectively.
Next we discuss the second term in eq.~(\ref{simpler-free-energy}).
Figure \ref{f:06} shows $\frac{1}{N^2} \log w_n(x)$
as a function of $x$ for $N=24$.
We find that the values at $x = 0.355$ 
for $n=3,4,5,6$ differ at most by 0.001.
Thus the results for 
${\cal F}_{{\rm SO}(d)}$ with $d=2,3,4,5$ are quite close to
each other.
While our results do not contradict
with an SO($3$) symmetric true vacuum 
as predicted by the GEM \cite{10070883},
it seems difficult to determine the true vacuum only 
from our data.

\section{Summary and discussions}
\label{sec:conclusions}
We have performed Monte Carlo studies of a Euclidean six-dimensional
version of the IIB matrix model, which is conjectured 
to be a nonperturbative definition of superstring theory.
This model was
studied before in the oneloop approximation,
where the computational effort is ${\cal O}(N^2)$ 
smaller \cite{0108041}. 
The simulation of the full model was made feasible by using 
state-of-the-art 
algorithms applied in recent studies of lattice QCD with
dynamical fermions. A strong complex-action problem makes calculations
very hard, and the factorization method was used in order to simulate
the system efficiently in the region favored by the competing effects
of the fluctuating phase, the real part of the action and the density
of states. The maximum and the saddle-points
of the distribution function for the
eigenvalues $\lambda_n$ of the tensor $T_{\mu\nu}$ were numerically
computed by solving the equations \rf{2.30},
and the VEVs
$\vev{\lambda_n}$ were estimated for SO($d$) symmetric vacua.
The large-$N$ scaling properties of the factorized distribution
functions 
allowed efficient extrapolations of the functions to large $N$,
which played an important role in the calculations.

The main goal of this work was to study, from first principles,
the scenario for the dynamical
compactification of space-time via SSB of 
SO($6$) 
yielding SO($d$) symmetric vacua with $d\le 5$.
The GEM calculations in ref.\cite{10070883} provided strong support 
for this scenario by showing that the SO(3)
symmetric vacuum has the lowest free energy.
Moreover, the GEM predicted 
that the SO($d$) symmetric vacua 
are characterized by a universal length scale $r$ for
the compactified dimensions,
and that the length scale $R_d$ for the large dimensions
satisfies $(R_d)^d r^{6-d}=\ell^6$
with some dynamical constant $\ell$, which
may be understood as the constant volume property of the system. 
Our results are consistent with
these predictions.
First we find that 
the eigenvalues $\vev{\lambda_n}_0$
in the phase-quenched model, which does not have SSB,
coincide in the large-$N$ limit 
with the value of $\ell^2$ obtained by the GEM.
Second we obtain a $d$ independent ($ 2 \le d \le 5$) value of
$r$ consistent with the numerical value given by the GEM.
Third our results for the large dimensions
are consistent with the constant volume property.
%
We have confirmed the
dramatic role of the fluctuations of the phase $\Gamma$ by showing
the absence of SSB in the phase-quenched model
and the effect of $w_n(x)$ in generating
dynamically the scale $r$ instead of $\ell$ of the phase-quenched
model. In actual determination of $r$, a crucial role is played by
the hard core potential generated 
by the nonperturbative short distance dynamics of
the eigenvalues, which was invisible
in the oneloop approximation used in 
ref.~\cite{0108041}. 

We have also attempted to compare the free energy for
SO($d$) symmetric vacua ($ 2 \le d \le 5$).
Although the comparison turned out to be subtle,
%
we feel that the overall consistency between
our results and the GEM calculations 
suggests that the conclusion obtained by 
the GEM \cite{10070883} is correct.
In the case of ten-dimensional IIB matrix model,
the GEM \cite{Nishimura:2011xy} 
predicts that SSB of SO($10$) also occurs, 
giving an SO($3$) symmetric vacuum with three large dimensions.
The physical interpretation of this statement requires special care,
however, especially in the light of the
recent results in ref.\cite{Kim:2011cr}, where a (3+1)-dimensional
{\it expanding} universe is 
found to arise dynamically in the {\it Lorentzian} IIB
matrix model.

Last but not the least, we consider that our analysis 
demonstrates how the factorization method can be 
used
in understanding nonperturbative dynamics
of a complex-action system 
by Monte Carlo simulations.
%
In particular, it is very encouraging that the method enabled
us to 
study
a physically interesting system 
in spite of the severe complex-action problem.
Since the basic idea is quite general, we hope that the method
can be applied to 
various
interesting systems which are
otherwise difficult to study.

\acknowledgments
We would like to thank
Tatsumi Aoyama, Masanori Hanada
and Shingo Takeuchi for collaborations at the early stage of
this work.
The authors are also grateful to Hajime Aoki, Satoshi Iso, 
Hikaru Kawai, Yoshihisa Kitazawa and Asato Tsuchiya 
for valuable discussions.
The work of T.A.\ and J.N.\ was
supported in part by Grant-in-Aid for Scientific Research
(No.\ 23740211 for T.A.\ and 20540286, 23244057 for J.N.) 
from Japan Society for the Promotion of Science.
This research has been co-financed by the European Union (European Social
Fund) and Greek national funds through the Operational
Program ``Education and Lifelong Learning'' of the National Strategic
Reference Framework Research Funding Program: ``THALES: Reinforcement
of the interdisciplinary and/or inter-institutional research and
innovation''. 

\appendix
 \section{Details of the Monte Carlo simulation}
 \label{sec:algorithm}
In this appendix we describe the details of the algorithm used in
the Monte Carlo simulations of the system \rf{3.1},
where $S_0$ is given by (\ref{2.7}).
First we define the matrix ${\cal D}= {\cal M}^\dag {\cal M}$ so that
$|\det{\cal M}|=\det {\cal D}^{1/2}$. In the Rational Hybrid Monte
Carlo (RHMC) algorithm \cite{Kennedy:1998cu}, 
we use the rational approximation
\begin{equation}
\label{a.6}
x^{-1/2} \simeq a_0 +\sum_{k=1}^Q\frac{a_k}{x+b_k}\  ,
\end{equation}
where $Q$ is chosen so that the error is small enough within the range
of $x$ required by the simulation. The real positive parameters $a_k$
and $b_k$  can be obtained by the code in ref.~\cite{remez} 
using the Remez algorithm. 
Then it is possible to approximate $\det{\cal D}^{1/2}$
as 
\begin{eqnarray}
\label{a.7}
\det{\cal D}^{1/2}&\simeq&\int \,  dF dF^* \, {\rm
  e}^{-S_{\mbox{\tiny PF}}[F,F^*,A]}\  , \\
\label{a.8}
\mbox{where~~~}
S_{\mbox{\scriptsize PF}}[F,F^*,A]&=& \tr\left\{
 a_0 F^\dagger F + 
 \sum_{k=1}^Q a_k F^\dagger \left( {\cal D}+b_k\right)^{-1} F
                         \right\}\  ,
\end{eqnarray}
and $(F_\alpha)_{ij}$ 
are auxiliary complex variables called
pseudofermions. The range of the spectrum of $\cal D$ determines the
accuracy goal in eq.~\rf{a.6}. 

In order to generate configurations, 
we use the Hybrid Monte Carlo method
with the ``Hamiltonian'' $H$, which evolves the system
in the fictitious time $\tau$, where
\begin{eqnarray}
\label{a.9}
H &=& \frac{1}{2}\tr \Pi^2 
+ \tr  \tilde\Pi^{\dagger} \tilde\Pi
+  S_{\mbox{\scriptsize eff}}[F,F^*,A]\  , \\
\label{a.10}
\mbox{and~~~}
S_{\mbox{\scriptsize eff}}[F,F^*,A] &=& 
 S_{\rm b}[A] + S_{\mbox{\scriptsize PF}}[F,F^*,A] + V(\lambda_n[A])\  .
\end{eqnarray}
The auxiliary variables 
$\tilde\Pi^\alpha_{ij}$ and 
$\Pi^\mu_{ij} = (\Pi^*)^\mu_{ji}$
are defined to be canonical momenta of the  
$((F_\alpha)_{ij},(A_\mu)_{ij})$ degrees of freedom so that
$\displaystyle\int\, d\tilde\Pi d\tilde\Pi^* d\Pi  dF dF^* dA\,{\rm
  e}^{-H} = \displaystyle\int\, dF dF^* dA\,{\rm
  e}^{-S_{\mbox{\tiny eff}}}$ and that the equations of motion 
\begin{eqnarray}
\label{a.11}
\frac{d A_\mu}{d\tau} =
 \alpha \frac{\partial H}{\partial \Pi^\mu} = \alpha\Pi^*_\mu
 \  ,& &\qquad
\frac{d F_\beta}{d\tau}  =
 \tilde\alpha \frac{\partial H}{\partial \tilde\Pi^\beta} = 
\tilde\alpha\tilde\Pi^*_\beta \  ,\nonumber\\
\frac{d \Pi^\mu}{d\tau} =
-\alpha \frac{\partial H}{\partial A_\mu} = 
-\alpha \frac{\partial S_{\mbox{\scriptsize eff}}}{\partial A_\mu}
 \  ,& &\qquad
\frac{d \tilde\Pi^\beta}{d\tau} =
-\tilde\alpha \frac{\partial H}{\partial F_\beta} = 
-\tilde\alpha \frac{\partial S_{\mbox{\scriptsize eff}}}{\partial F_\beta}
\  ,
\end{eqnarray}
for the evolution in the fictitious time $\tau$ preserve the
Hamiltonian $H$. 
The real-positive coefficients $\alpha$, $\tilde\alpha$ are Fourier
acceleration coefficients \cite{Catterall:2001jg}, which, if optimized,
can greatly reduce autocorrelation times. In our case we use
$\alpha^2 \simeq \dfrac{1}{6 N^2} \vev{\tr (A_\mu)^2}$ and
$\tilde\alpha^2 \simeq 
\dfrac{1}{4 N^2} \vev{\tr (F_\beta)^2}$.

This fictitious time-evolution is called the Molecular Dynamics.
Numerically it is implemented by discretizing the equations of
motion \rf{a.11} with a time step 
$\Delta\tau$ and then 
integrating from $\tau=0$ to $\tau=\tau_{\rm f}$ in
$N_\tau$ steps, so that $\tau_{\rm f}=N_\tau\,\times\Delta\tau$.
For that, the so-called leapfrog discretization is used 
in order to maintain time reversibility. 
The Hamiltonian $H$ is now not conserved due to discretization
errors and $\Delta H = H(\tau_{\rm f})-H(0)$ is of ${\cal O}(\Delta\tau^2)$.
A Metropolis
accept/reject decision  maintains detailed balance in order to obtain
the correct distribution in the sampled configurations.  
The updating procedure
therefore consists of the following steps:
\begin{itemize}
\item Refresh the momenta $(\tilde\Pi^\alpha(0),\Pi^\mu(0))$ according
  to their Gaussian 
${\rm e}^{-\mbox{\scriptsize tr} \tilde\Pi^{\dagger} \tilde\Pi }$, 
${\rm  e}^{-\frac{1}{2}\mbox{\scriptsize tr} \Pi^2}$ distributions. 
This is
  necessary and sufficient for maintaining ergodicity.
\item Evolve the $(\tilde\Pi^\alpha(0),\Pi^\mu(0),F_\alpha(0),A_\mu(0))$
  configuration for time $\tau_{\rm f} = N_\tau \times\Delta\tau$ using the
  discretized versions of eq.~\rf{a.11}. 
\item Accept or reject the $(F_\alpha(\tau_{\rm f}),A_\mu(\tau_{\rm f}))$ configuration with
  probability $\mbox{min}(1,{\rm e}^{-\Delta H})$,
where $\Delta H=H(\tau_{\rm f})-H(0)$.
\end{itemize}
In the process, we have to keep $\tau_{\rm f}$ 
large enough in order to reduce
autocorrelation times and $\Delta\tau$ small enough in order to
maintain reasonable acceptance rates. 
In practice, we first fix $\tau_{\rm f}$
and optimize $\Delta\tau$ by maximizing
$\Delta\tau\times(\mbox{acceptance rate})$. 
Then we optimize $\tau_{\rm f}$
by minimizing autocorrelation times in units of 
Molecular Dynamics step. 
For more details, see ref.~\cite{Ambjorn:2000dx}.

The main part of our computational effort is spent in the calculation
of the terms containing $({\cal D}+b_k)^{-1} F$,
which appear in
the Molecular Dynamics integration steps, as well as in the
calculation of the Hamiltonian $H$. These terms are replaced by the
solutions $(\chi_k)_{\alpha,ij}$ to the linear system $({\cal D}+b_k)
\chi_k = F$. For this we employ the conjugate gradient method for the
smallest of the coefficients $b_k$ and then use multimass Krylov
solvers to compute the solutions for the other $b_k$
\cite{Jegerlehner:1996pm}. This way we avoid ${\cal O}(Q)$ of
arithmetic operations. The conjugate gradient method requires 
${\cal O}(N^3)$
arithmetic operations\footnote{Naively
one needs 
${\cal  O}(N^4)$ 
operations, but one notes that
the nonzero elements in $\cal M$ are 
${\cal  O}(N^3)$.
In actual calculation, we 
perform
the multiplication of $\chi_k$ 
by $\cal M$ 
using the expression
$\Gamma_\mu [A_\mu , \chi_k] $ as in ref.~\cite{Ambjorn:2000bf}.} 
for each multiplication of $\chi_k$ by ${\cal M}$ \cite{Ambjorn:2000bf}. 
The number of iterations for the convergence of the method is 
${\cal  O}(N^2)$, 
so each Molecular Dynamics step costs 
${\cal O}(N^5)$. 
This is in contrast to
the QCD case, where the conjugate gradient method converges after a
number of iterations which is independent of the system size.


The term $\dfrac{\partial V(\lambda_n[A])}{\partial A_\mu}$,
which appears in the left-bottom equation in (\ref{a.11}),
is calculated as follows. Using the relation
\begin{equation}
\label{a.12}
 \sum_{\rho=1}^{6} T_{\nu \rho} v_{\rho}^{(n)} 
= \lambda_{n} v_{\nu}^{(n)},
\end{equation}
where $v_{\mu}^{(n)}$ ($n=1,\ldots,6$) 
are the six eigenvectors of the $6 \times 6$ matrix
$T_{\mu \nu}=\dfrac{1}{N}\tr (A_\mu A_\nu)$ normalized as $\sum_{\mu=1}^{6}
v^{(n)}_{\mu} v^{(n)}_{\mu} = 1$ for each $n$. Taking the derivative of
eq.~\rf{a.12} with respect to $(A_{\mu})_{kl}$, we obtain
\begin{equation}
\label{a.13}
 \sum_{\rho=1}^{6} 
\left(  \frac{\partial T_{\nu \rho}}{\partial (A_{\mu})_{kl}}
v_{\rho}^{(n)} + 
T_{\nu \rho} \frac{\partial v_{\rho}^{(n)}}{\partial (A_{\mu})_{kl}} 
\right)
 =  \frac{\partial \lambda_{n}}{\partial (A_{\mu})_{kl} } 
v_{\nu}^{(n)} + \lambda_{n} 
\frac{\partial v^{(n)}_{\nu}}{\partial (A_{\mu})_{kl} }  \ .
\end{equation}
Multiplying both sides 
of eq.~\rf{a.13}
by $v_{\nu}^{(n)}$ and taking a sum over $\nu$, we obtain
\begin{equation}
\label{a.14}
\sum_{\nu,\rho=1}^{6} v_{\nu}^{(n)} 
 \frac{\partial T_{\nu \rho}}{\partial (A_{\mu})_{kl}} 
 v_{\rho}^{(n)}  = 
\sum_{\nu=1}^{6} v_{\nu}^{(n)} 
 \frac{\partial \lambda_{n}}{\partial (A_{\mu})_{kl} } 
 v_{\nu}^{(n)} = 
 \frac{\partial \lambda_{n}}{\partial (A_{\mu})_{kl} }  \ ,
\end{equation}
where the second terms of each side of eq.~\rf{a.13} cancel. 
Therefore we obtain
\begin{equation}
\label{a.15}
\frac{\partial V(\lambda_{n})}{\partial (A_{\mu})_{kl}} 
= \gamma_{n} (\lambda_{n} - \xi_{n}) 
\frac{\partial \lambda_{n}}{\partial (A_{\mu})_{kl}} 
= \frac{2 \gamma_{n}}{N} (\lambda_{n} - \xi_{n}) 
\sum_{\nu=1}^{6} v^{(n)}_{\mu} v^{(n)}_{\nu} (A_{\nu})_{lk} \ .
\end{equation}

In order to calculate the phase of the fermion determinant, we
calculate ${\rm det}{\cal M}[A]$ using (\ref{def-Zf})
explicitly as described in ref.~\cite{Ambjorn:2000bf}.
We define a complete basis of the
general complex $N\times N$ matrices $t^a \in gl(N,\mathbb{C})$ by
$(t^a)_{ij} = \delta_{i i_a} \delta_{j j_a}\  $,
where 
$a=1,\ldots,N^2$, $i_a, j_a = 1,\ldots, N$ and 
$a=N(i_a-1)+j_a$. 
Taking into account the tracelessness of the fermionic
matrices, the integration of $\psi$, $\bar\psi$ yields 
$\det{\cal M}[A]$,
where the $4 (N^2-1)\times 4 (N^2-1)$ matrix 
${\cal M}[A]$
is given by \cite{Ambjorn:2000bf}
\begin{equation}
\label{a.4}
   {\cal M}_{a \alpha, b \beta} = {\cal M}'_{a \alpha, b \beta}
   - {\cal M}'_{N^{2} \alpha, b \beta} \, \delta_{i_{a} j_{a}} 
   - {\cal M}'_{a \alpha, N^{2} \beta} \, \delta_{i_{b} j_{b}} 
\ ,
\end{equation}
with $\alpha,\beta=1,\ldots, 4$ and
the $4 N^2\times 4 N^2$ matrix ${\cal M}'$ defined by
\begin{equation}
\label{a.5}
   {\cal M}'_{a \alpha, b \beta} =
   (\Gamma_{\mu})_{\alpha \beta} \, \tr \, ( t^{a} [A_{\mu}, t^{b}]) \ . 
\end{equation}



\end{document}